\documentclass[preprint,authoryear,12pt]{elsarticle0}

\usepackage{amssymb}
\usepackage{subfigure}

\usepackage{multirow}
\usepackage{colortbl}

\begin{document}

\begin{frontmatter}



\title{An MRI-Derived Definition of MCI-to-AD Conversion for Long-Term, Automatic Prognosis of MCI Patients\\
{\normalsize for the Alzheimer's Disease Neuroimaging 
Initiative}\tnoteref{label1}}
\tnotetext[label1]{Data 
used in preparation of this article were obtained from the Alzheimer's Disease Neuroimaging Initiative (ADNI) database (adni.loni.ucla.edu). As such, the investigators within the ADNI contributed to the design and implementation of ADNI and/or provided data but did not participate in analysis or writing of this report. A complete listing of ADNI investigators can be found at: http://adni.loni.ucla.edu/wp-content/uploads/how\_to\_apply/ADNI\_Authorship\_List.pdf
}



\author[CNMRR]{Yaman Aksu}
\author[EE]{David J. Miller}
\author[EE,CSE]{George Kesidis}
\author[CENI]{Don C. Bigler}
\author[CNMRR]{Qing X. Yang}
\address[CNMRR]{Center for NMR Research, Department of Radiology, Penn State University College of Medicine, Hershey, PA, USA}
\address[EE]{Electrical Engineering Department, Penn State University, University Park, PA, USA}
\address[CSE]{Computer Science and Engineering Department, Penn State University, University Park, PA, USA}
\address[CENI]{Center for Emerging Neurotechnology and Imaging, Penn State University College of Medicine, Hershey, PA, USA}
\begin{abstract}
Alzheimer's disease (AD), and its precursor state, mild cognitive
impairment (MCI), continue to be widely studied. While there is no
consensus on whether MCIs actually ``convert'' to AD, this {\it
concept} is widely applied, allowing statistical testing and machine
learning methods to help identify early disease biomarkers and build
models for predicting disease progression. Thus, the more important
question is not {\it whether} MCIs convert, but {\it what is the
best such definition}. We focus on automatic prognostication,
nominally using only a baseline image brain scan, of whether an MCI
individual will convert to AD within a multi-year period following
the initial clinical visit. This is in fact not a traditional
supervised learning problem since, in ADNI, {\it there are no
definitive labeled examples of MCI conversion}. It is not truly
unsupervised, either, since there are (labeled) AD and Control
subjects, as well as clinical and cognitive scores for MCIs. Prior
works have defined MCI subclasses based on whether or not
clinical/cognitive scores significantly change from baseline. There
are serious concerns with these definitions, however, since e.g.
most MCIs ({\it and ADs}) do not change from a baseline CDR=0.5 at
any subsequent visit in ADNI, even while physiological changes may
be occurring. These works ignore rich {\it phenotypical} information
in an MCI patient's brain scan and labeled AD and Control examples,
in defining conversion. We propose an innovative conversion
definition, wherein an MCI patient is declared to be a converter if
any of the patient's brain scans (at follow-up visits) are
classified ``AD'' by an (accurately-designed) Control-AD classifier.
This novel definition {\it bootstraps} the design of a {\it second}
classifier, {\it specifically trained to predict whether or not MCIs
will convert}. This second classifier thus predicts whether an
AD-Control classifier will predict that a patient has AD. Our
results demonstrate this new definition leads not only to much
higher prognostic accuracy than by-CDR conversion, but also to
subpopulations much more consistent with {\it known} AD brain region
biomarkers. We also identify key {\it prognostic} region biomarkers,
essential for accurately discriminating the converter and
nonconverter groups.
\end{abstract}


\end{frontmatter}


\section{Introduction} \label{sec:ad-intro}
The dementing illness Alzheimer's disease (AD), and the transitional
state between normal aging and AD referred to as mild cognitive
impairment (MCI) continue to be widely studied. Individuals
diagnosed with MCI have memory impairment, yet without meeting
dementia criteria. Annually $\approx$ 10-15\% of people with MCI are
diagnosed with AD \citep{Petersen_2004}. Moreover, prior to symptom
onset, brain abnormalities have been found in people with MCI, as
ascertained by retroactive evaluation of longitudinal MRI scans
\citep{Davatzikos_2008}. There is no consensus on whether MCI
patients actually ``convert'' to AD. First, some MCI patients may
suffer from other neurodegenerative disorders (e.g., Lewy body
dementia, vascular dementia and/or frontotemporal dementia). Second,
it is possible that all other MCI patients already have AD, but at a
preclinical stage. AD diagnosis itself may not be considered
definitive without e.g. confirming pathologies such as the amyloid
deposits detectable at autopsy. Regardless of whether MCI patients
truly ``convert'' to AD or not, the concept of MCI-to-AD conversion
has been widely applied, e.g. \citep{Chou_SPIE, Davatzikos_2010,
Misra_2009, Vemuri_2009, DZhang_2011} and is utilitarian -- defining
MCI (converter and nonconverter) subgroups allows use of statistical
group difference tests and machine learning methods to help identify
early disease biomarkers and to build models for predicting disease
progression. For these purposes, the more important question is not
whether MCIs convert, but rather {\it what is the best such
definition}.

Accordingly, here we focus on the following \textbf{Aim}: automatic
prognostication, (nominally) using only a baseline brain scan, of
whether an MCI individual will convert to AD within a multi-year
(three year) period following an initial (baseline) clinical
visit\footnote{Our system performs three-year ahead prediction
because it is designed based on the ADNI database, which followed
participants for a period of up to three years.}.

While only image voxel-based features are evaluated
here for use by our classifier, our framework is extensible to
incorporating other baseline clinical information (e.g. weight,
gender, education level, genetic information, and clinical cognitive scores such as the
Mini Mental State Exam (MMSE)) into the decisionmaking.  Moreover,
our approach can also incorporate the recent, promising
cerebrospinal fluid (CSF) based markers \citep{DeMeyer_2010}.
However, as this requires an invasive spinal tap, we focus here on
image scans, which are routinely performed for subjects with MCI.

We do not hypothesize that, within ADNI, there are actually two
subclasses of MCI subjects when evaluated over the very long term --
those that (eventually) convert to AD, and those that do not.  Even
if an overwhelming majority of MCI subjects will {\it eventually}
convert, identifying the subgroup likely to convert {\it within
several years} has several compelling utilities: 1) early prognosis,
to assist family planning; 2) facilitating group-targeted treatments/drug trials;
3) we identify key {\it prognostic} brain ``biomarker''
regions, {\it i.e.} those found to be most critical for accurately
discriminating our ``converter'' and ``nonconverter'' groups. These
regions may shed light on disease etiology.

Distinguishing AD converters from nonconverters is a binary
(two-class) classification problem. Moreover, it may appear this
classification problem can be directly addressed via supervised
learning methods \citep{Duda}. However, it is in fact an {\it
unconventional} problem, lying somewhere {\it between} supervised
classification and unsupervised classification (clustering), and
thus requiring a unique approach. To appreciate this, consider the
ADNI cohort of MCI individuals. ADNI consists of clinical
information and image scans on hundreds of participants, taken at
six-month intervals for up to three years. A clinical label (AD,
MCI, or Control) was assigned to each participant at first
visit\footnote{Clinicians derive the AD/MCI/Control label based on
multiple criteria, which may include Clinical Dementia Rating (CDR),
whose possible values are: {\it 0=none, 0.5=questionable, 1=mild,
2=moderate, 3=severe}.}. Even though a {\it probable AD} definition
based on CDR and MMSE scores and NINCDS/ADRDA criteria has been
used, e.g. \citep{Leow_2009, DZhang_2011}, to provide follow-up
assessment for MCI patients, this is strictly a clinically driven
definition, based on a clinical rating (CDR) and a cognitive score
(MMSE) whose difficulties will be pointed out shortly. This is not a
definitive (autopsy-based) determination of AD, nor is it a
definition based on physiological brain changes. Even if the {\it
probable AD} definition has very high {\it specificity}, it may not
be sufficiently sensitive, i.e. there may be patients who are
undergoing significant physiological brain changes consistent with
conversion, yet without clinical manifestation.

Accordingly, we will approach
the conversion problem from a perspective as agnostic and unbiased
as possible, and simply state that it is not definitively known
which MCI participants in ADNI truly converted to AD within three
years. In conventional supervised classifier learning, one has
labeled training examples, used for designing the classifier, and
labeled test examples, used to estimate the classifier's
generalization accuracy. For predicting whether MCI participants in
ADNI convert to AD,  we in fact have neither. Thus, our problem is
not conventional supervised learning. On the other hand, consider
{\it unsupervised} clustering \citep{Duda}.  Here, even if one knows
the number of clusters (classes) present, there is no prior
knowledge on what is a good clustering -- one is simply looking for
underlying grouping tendency in the data.  Clearly, our problem does
not fit unsupervised clustering, either -- while we have no labeled
MCI converter/nonconverter instances per se, 1) there are two designated classes of
interest (converter and nonconverter); and 2) there are known class
characteristics -- conversion to AD should, plausibly, mean
that: (i) a clinical measure such as CDR or a cognitive measure has
changed {\it and/or} (ii) there are changes in brain features more
characteristic of AD subjects than normal/healthy subjects. Note
that in ADNI we do have plentiful labeled AD and normal/healthy
(Control) examples to help assess ii).

Based on the above, MCI prognosis is an interesting and novel
problem, lying somewhere between supervised and unsupervised
classification. The crux of this problem is to craft criteria through
which meaningful MCI subgroups can be defined, well-capturing
notions of ``AD converter'' and ``nonconverter''. To help guide
development and evaluation of candidate definitions, we state the
following three {\bf desiderata}: 1) The proposed definition of AD
converter should be plausible and should exploit the available,
relevant information in the ADNI database (e.g. image data, labeled
AD and Control examples, and clinical information). To appreciate
1), note that the MCI population could be dichotomized in {\it many}
ways, {\it e.g.} by height, and there might be significant
clustering tendency with respect to height, but such a grouping is
likely meaningless for MCI prognosis; 2) A classifier trained based
on these class definitions should {\it generalize} well on test data
(not used for training the classifier) -- this quantifies how accurately we can discriminate the classes
that we have defined. If we create what we believe to be good
definitions, but ones that cannot be accurately discriminated, that
would not be useful clinically; 3) The class definitions should be
validated using known AD conversion biomarkers ({\it i.e.}, external
measures) such as measured changes from baseline both in volumes of
brain regions known to be associated with the disease
\citep{Schuff_2010} and in cognitive test scores (such as the
clinical MMSE measure).

\noindent {\bf Prior Related Work}

Several prior works, e.g. \citep{Davatzikos_2010, Misra_2009,
Wang_2010}, defined converter and nonconverter classes {\it solely}
according to whether the baseline visit CDR score of $0.5$ rose or
stayed the same over all visits. Change in CDR has also been used as
surrogate ground-truth for cognitive decline in a number of other
papers, e.g. \citep{Chou_ADNI, Vemuri_2009}. While CDR gives a
workable conversion definition, it should be evaluated with respect
to the three desiderata above. We will evaluate 2) and 3) in the
sequel. With respect to 1), one should challenge a CDR-based
conversion definition. First, CDR is not an effective discriminator
between the AD and MCI groups, {\it i.e.} there is {\it very
significant} AD-MCI overlap, not only with respect to CDR=1 {\it but
even 0.5} -- for ADNI, the majority of the hundreds of AD subjects
used in our experiments start (at first visit) at CDR=0.5 and stay
at 0.5 at {\it all} later visits; likewise, nearly all MCI subjects
start at 0.5, with a large majority of these {\it also} staying at
0.5 for all visits. This latter fact further implies difficulties in
finding an adequate number of conversion-by-CDR subjects in ADNI,
both for accurate classifier training and test set evaluation.  For
the even more stringent {\it probable AD} definition (meeting MMSE
and NINCDS/ADRDA criteria, in addition to CDR changing from 0.5 to
1) there are necessarily even {\it fewer} MCI converters for
classifier training and testing. Second, a purely CDR-based (or
``probable AD" based) conversion definition ignores the (rich) {\it
phenotypical} information in an MCI subject's image brain scans and
does not exploit the labeled AD and normal/healthy (Control)
examples in ADNI. These prior works do treat features derived from
brain scans as the {\it covariates} (the inputs) to the
classifier/predictor. However, we believe the MCI brain scans can
themselves be used, in conjunction with the labeled AD and Control
examples, to help define more accurate surrogate ground-truth.

Previous work has demonstrated that structural MRI analysis is
useful for identifying AD biomarkers in individual brain regions
\citep{Chetelat_2002, Fan_2008, Fennema_2009} -- {\it e.g.},
cortical thinning \citep{Lerch_2005, Thompson_2003}, ventricle
dilation and gaping \citep{Chou_ADNI, Chou_SPIE, Schott_2005},
volumetric and shape changes in the hippocampus and entorhinal
cortex \citep{Csernansky_2005, DeLeon_2006, Stoub_2005}, and
temporal lobe shrinkage \citep{Rusinek_2004}. It is important to
capture interaction effects across multiple brain regions.
\citep{Davatzikos_2008, Misra_2009, Vemuri_2008, Wang_2010} did
jointly analyze voxels (or regions) spanning the entire brain and
did build classifiers or predictors. Moreover, as part of their
work, \citep{Wang_2010} investigated prediction of future decline in
MCI subjects working from baseline MRI scans, which is the primary
subject of our current paper. However, there are several limitations
of these past works. First, all these studies used the previously
discussed CDR and cognitive measures such as MMSE, which has been
described as noisy and unreliable, as the ground-truth prediction
targets for classifier/regressor training. In \citep{Chou_ADNI}, the
authors state: ``Cognitive assessments are notoriously variable over
time, and there is increasing evidence that neuroimaging may provide
accurate, reproducible measures of brain atrophy.'' Even in
\citep{Wang_2010}, where MMSE was treated as the measure of decline
and the ground-truth regression target, the authors acknowledged
that ``individual cognitive evaluations are known to be extremely
unstable and depend on a number of factors unrelated to...brain
pathology.'' Such factors include sleep deprivation, depression,
other medical conditions, and medications. Even though MMSE is
widely used by clinicians, these comments (even if not universally
accepted), do indicate MMSE by itself may not be so reliable in
quantifying the disease state. Moreover, while \citep{Wang_2010} did
build predictors of future MMSE scores working from baseline scans,
this was not a main focus of their paper -- their paper focused on
predicting the current score.  Their prognostic experiments involved
a very small sample size (just 26 participants from the ADNI
database). Accordingly, it is difficult to draw definitive
conclusions about the accuracy of their prognostic model and their
associated brain biomarkers. The main reason the authors chose such
a small sample was, as the authors state: ``A large part
of...ADNI...are from patients who did not display significant
cognitive decline...[these] would overwhelm the regression algorithm
if..used in the...experiment.''  While this statement (with
cognitive decline measured according to MMSE) may be true, that does
not mean many of those excluded ADNI subjects are not experiencing
significant physiological brain changes/atrophy.  The novel approach
we next sketch is well-suited to identifying MCI subjects undergoing
such changes.

\noindent {\bf Our Neuroimaging-Driven, Trajectory-Based Approach}

Here, we propose a novel approach for prognosticating putative
conversion to AD driven by image-based information (and exploiting
AD-Control examples), rather than by a single, non-image-based,
weakly discriminating clinical measurement such as CDR. Our solution
strategy is as follows. We first build an accurate image-based
Control-AD classifier ({\it i.e.}, using AD and Control subjects, we
build a Support Vector Machine (SVM) classifier) (Vapnik, 1998). We
then apply this classifier to a {\it training population} of MCI
subjects -- separately, for each subject visit, we determine whether
the subject's image is on the AD side or the Control side of the
SVM's fixed (hyperplane) decision boundary.  In addition to a binary
decision, the SVM gives a ``score'' -- essentially the distance to
the classifier's decision boundary.  Thus, for each MCI subject, as a
function of visits, we get an image-based ``phenotypical'' score
trajectory. We fit a line to each subject's trajectory and extend
the line to the sixth visit if the sixth visit is missing.  We can
then give the following {\it trajectory-based conversion
definition}: if the extended line either starts on the AD side or
crosses to the AD side over the six visits, we declare this person a
``converter-by-trajectory''. Otherwise, this person is a
``nonconverter-by-trajectory''\footnote{A very small percentage of
the MCI population, in our experiments often $1\%$ and not exceeding
$5\%$, may unexpectedly start on the AD side and cross to the
Control side. We treat these individuals as outliers and omit them
from our experiments.}. In this fashion, we {\it derive}
ground-truth ``converter'' and ``nonconverter'' labels for an
(initially unlabeled) training MCI population. These (now) labeled
training samples {\it bootstrap} the design of a {\it second} SVM
classifier which uses only the first-visit training set MCI images
and is trained to predict whether or not an MCI patient is a
``converter-by-trajectory''. Essentially, this second (prognostic)
classifier is predicting whether, within three years, an AD-Control classifier will
predict that a patient has AD. Via these two classifier design
steps, we thus build a classification system for our
(unconventional) pattern recognition task.

SVMs are widely used classifiers whose accuracy is attributed to
their maximization of the ``margin'', {\it i.e.} the smallest
distance from any training point to the classification boundary.
Since the SVM finds a linear discriminant function that {\it
maximizes} margin, a significant change in score is generally needed
to cross from the control side to the AD side, which is thus
suggestive of conversion from MCI to AD.  This is the premise
underlying our approach.

The main contributions of our work are: 1) a novel image-based
prognosticator of MCI-to-AD conversion that we will demonstrate to
achieve both better generalization accuracy and much higher
correlation with known brain region biomarkers than the CDR-based
approach; 2) The finding that MCI subgroups that are strongly
correlated with known AD brain region biomarkers are not so strongly
correlated with ``cognitive decline'' as measured by MMSE; 3)
Identification of the brain regions most critical for accurately discriminating
between our ``converter'' and ``nonconverter'' groups, via
application of margin-based feature selection (MFE) \citep{Aksu_TNN}
to brain image classification, and demonstration of MFE's better
performance than the well-known RFE method \citep{Guyon_RFE} on this
domain. 
\section{Methods} \label{sec:methods}
\subsection{Subjects and MRI data} \label{sec:mri}
We used $T_{1}$-weighted ADNI images\footnote{
Data used in the preparation of this article were obtained from the Alzheimer's Disease Neuroimaging Initiative (ADNI) database (adni.loni.ucla.edu). The ADNI was launched in 2003 by the National Institute on Aging (NIA), the National Institute of Biomedical Imaging and Bioengineering (NIBIB), the Food and Drug Administration (FDA), private pharmaceutical companies and non-profit organizations, as a \$60 million, 5-year public- private partnership. The primary goal of ADNI has been to test whether serial magnetic resonance imaging (MRI), positron emission tomography (PET), other biological markers, and clinical and neuropsychological assessment can be combined to measure the progression of mild cognitive impairment (MCI) and early Alzheimer's disease (AD). Determination of sensitive and specific markers of very early AD progression is intended to aid researchers and clinicians to develop new treatments and monitor their effectiveness, as well as lessen the time and cost of clinical trials.
}
that have undergone image
correction described at the ADNI website.\footnote{ADNI image
correction steps include Gradwarp, N3, and scaling for gradient
drift -- see www.loni.ucla.edu/ADNI/Data/ADNI\_Data.shtml.} ADNI
aims to recruit and follow 800 research participants in the 55-90
age range: approximately 200 elderly Controls, 400 people with MCI,
and 200 people with AD. The number of Control, MCI, and AD
participants in our analysis were $\approx$180, 300, and 120,
respectively -- experiment-specific detailed descriptions will be
provided in Sec. \ref{sec:results}. We processed the
$T_{1}$-weighted images as described in \ref{sec:image-proc},
producing new images from which we then obtained the features (next
discussed) used by our statistical classifiers.
\subsection{Features for classification} \label{sec:features}
We chose as features the voxel intensities of a
processed\footnote{We describe our processing of RAVENS images in
\ref{sec:image-proc}.} RAVENS image, a type of ``volumetric
density'' image \citep{Davatzikos_1998, Davatzikos_2001,
Goldszal_1998, Shen_2003} that has been validated for voxel-based
analysis \citep{Davatzikos_2001} and applied both to AD e.g.
\citep{Davatzikos_2010, Misra_2009, Wang_2010} and other studies
e.g. \citep{Fan_COMPARE}\footnote{Of particular interest,
\citep{Davatzikos_2001} supported that voxel-based SPM statistical
analysis, which we perform herein for comparison with our methods,
can be performed on RAVENS images.}.
For each of the three processed RAVENS tissue maps (gray matter
(GM), white matter (WM), and ventricle), to reduce complexity for
subsequent processing, we obtained a subsample by successively
skipping five voxels along each of the three dimensions, and took as
feature set the union of the three subsampled maps. We will
also report results for the case of skipping only two voxels,
rather than five.

Since high-dimensional
nonlinear registration (warping) of all individuals to a common
atlas (via HAMMER \citep{Shen_HAMMER_ITMI}) is applied in producing
our features, they 
capture both volumetric and morphometric brain characteristics,
which is important since individuals with AD/MCI typically exhibit
brain atrophy (affecting both volume and shape).
\subsection{Classification and feature selection for high-dimensional
images} \label{sec:mfe-intro} A challenge in building classifiers
for medical images is the relative paucity of available training
samples, compared to the huge dimensionality of the voxel space and,
thus, to the number of parameters in the classifier model -- in
general, the number of parameters may grow at least linearly with
dimensionality. In the case of 3D images, this could imply even
millions of parameter values (e.g. one per voxel) need to be
determined, based on a training set of only a few hundred patient
examples. In such cases, classifier overfitting is likely,
which can degrade generalization (test set) accuracy. Here we will
apply a linear discriminant function (LDF) classifier with a
built-in mechanism to avoid overfitting and with design complexity
that scales well with increasing dimensionality - the support vector
machine (SVM) \citep{Vapnik_1998}. The choice of LDF achieving
perfect separation (no classification errors) for a given two-class
training set is not unique. The SVM, however, is the unique
separating LDF that maximizes the {\it margin}, {\it i.e.} the minimum distance to
the classifier decision boundary, over all training samples. In this
sense, the SVM maximizes separation of the two classes. For an SVM,
unlike a standard LDF, the number of model parameters is bounded by
the number of training samples, rather than being controlled by the
feature dimensionality. Since the number of samples is the much
smaller number for medical image domains, in this way the SVM
greatly mitigates overfitting. SVMs have achieved excellent
classification accuracy for numerous scientific and engineering
domains, including medical image analysis, \citep{Aksu_TNN,
Davatzikos_2010, Guyon_RFE}.

Even though SVMs are effective at mitigating overfitting,
generalization accuracy may still be improved in some cases by
removing features that contribute little discrimination power.
Moreover, even if generalization accuracy monotonically improves
with increasing feature dimensionality, high complexity (both
computation and memory storage) of both classifier design and class
decisionmaking may outweigh small gains in accuracy achieved by
using a huge number of features. Most importantly here, it is often
useful to identify the critical subset of features necessary for
achieving accurate classification -- these ``markers'' may shed
light on the underlying disease mechanism. In our case, this will
help to identify prognostic brain regions, associated with MCI
conversion.

Unfortunately, there is a huge number of possible feature subsets,
with exhaustive subset evaluation practically prohibited even for
modest number of features, $M$, let alone $M \sim10^6$. Practical feature selection
techniques are thus heuristic, with a large range of tradeoffs
between accuracy and complexity \citep{Guyon_2003}. ``Front-end''
(or ``filtering'') methods select features prior to classifier
training, based on evaluation of discrimination power for individual
features or small feature groups. ``Wrapper'' methods are generally
more reliable, interspersing sequential feature selection and
classifier design steps, with features sequentially selected to
maximize the current subset's joint discrimination power. There are
also embedded feature selection methods, e.g. for SVMs, use of
$\ell_1$-regularization within the SVM design optimization
\citep{Fung-NLPSVM}, in order to find ``sparse'' weight vector
solutions, which effectively eliminate many features. For wrappers,
there is greedy forward selection, with ``informative'' features
added, backward elimination, which starts from the full set and
removes features, and more complex bidirectional searches. In our
work, due to the high feature dimension, we focus on two backward
elimination wrappers that afford practical complexity: i) the widely
used recursive feature elimination algorithm (RFE)
\citep{Guyon_RFE}, where at each step one removes the feature with
least weight magnitude in the SVM solution. RFE has been applied
before to AD \citep{Davatzikos_2010, Misra_2009, Wang_2010}; ii) the
recent margin-based feature elimination (MFE) algorithm
\citep{Aksu_TNN}, which uses the same objective function (margin)
for feature elimination, one consistent with good generalization,
that the SVM uses for classifier training \citep{Vapnik_1998}. MFE
was shown in \citep{Aksu_TNN} to outperform RFE \citep{Guyon_RFE}
and to achieve results comparable to embedded feature selection for
domains with up to 8,000 features (gene microarray classification).
Here we will also find that MFE gives better results than RFE.
\subsection{An MRI-Derived Alternative to CDR-based MCI-to-AD Conversion}
\label{sec:in-tandem} In the Introduction, we outlined our two
classifier design steps for building an automatic prognosticator for an
individual with MCI. In this section, we elaborate on these two
steps and give an illustrative example. Our AD-Control classifier,
used in the first step, is discussed in Sec.
\ref{sec:classifier1}, and our second classifier, used to
discriminate converter-by-trajectory (CT) and
nonconverter-by-trajectory (NT) classes, is discussed in Sec.
\ref{sec:classifier2}.
\subsubsection{AD-Control classifier} \label{sec:classifier1}
For the AD training population, we chose individual AD visit images
with a CDR score of at least 1.  For the Control training
population, on the other hand, we only chose {\it initial} visits,
and only those for participants who stayed at CDR=0 throughout all
their visits.
Thus, we excluded Controls with ``questionable
dementia'' (i.e., CDR=0.5) at any visit. By these choices, we sought
to exclude outlier examples or even possibly any mislabeled
examples, recalling that CDR for the majority of both AD and MCI
participants is 0.5 throughout all visits.
\subsubsection{CT-NT classifier} \label{sec:classifier2}
\begin{figure*}
\centering
\subfigure[]{\label{fig:extended-SH}\includegraphics[scale=0.35]{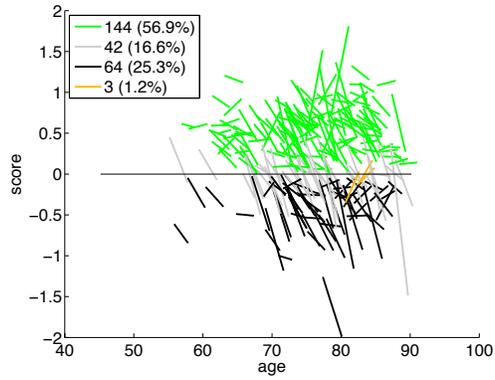}}
\subfigure[]{\label{fig:extended-RH}\includegraphics[scale=0.35]{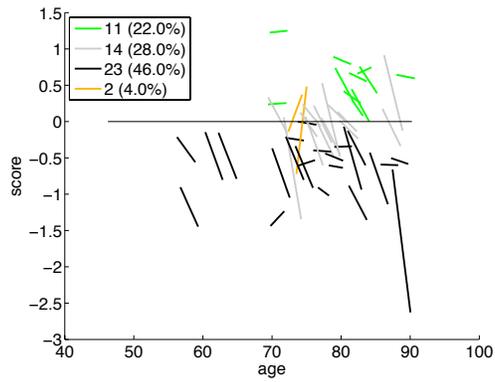}}
\caption{AD-Control SVM score trajectories for MCI subjects. (a)
Nonconverters-by-CDR. (b) Converters-by-CDR.}
\label{fig:traj-extended}
\end{figure*}
Fig. \ref{fig:traj-extended} gives an illustrative example of the
phenotypical score trajectories for MCI subject described in the Introduction. A
positive score is on the Control side and a negative score is on the
AD side -- the x-axis represents
the AD-Control SVM's decision boundary. Score vs. age is plotted, with each line
segment a trajectory obtained by linearly fitting an individual's
phenotype scores (and linearly extrapolating if there are missing
visits). Nonconverters-by-CDR (N-CDR) and converters-by-CDR (C-CDR)
are illustrated in (a) and (b), respectively. Green and black
subjects are those whose fitted trajectory stayed on the Control
side and AD side, respectively, whereas gray lines are subjects who
crossed to the AD side. Thus, by our conversion-by-trajectory
definition, the green group is the nonconverters-by-trajectory, and
the black and gray groups together are the converters-by-trajectory.
Subject counts for these groups are given in the figure legends. The
outlier subjects are shown in orange -- there are five, making up
less than $2\%$ of the MCI cohort. 
Notice, intriguingly, from the left figure that more than one third of all 
(non-orange) MCI patients (106 of 298) are converters by trajectory and yet nonconverters 
according to CDR -- {\it i.e.}, there is a very large percentage of patients for which the two converter 
definitions disagree, with the neuroimage-based definition indicating 
disease state changes that are not predicted using the clinical, CDR-based definition.
Likewise, an
additional $3\%$ of all MCI subjects (11 of 298) ``defy'' their by-CDR
converter label in that
they do not reach the AD side of the decision boundary.

Based on these trajectories, i.e. whether or not the AD side is
visited, we derive the ground-truth `CT' and `NT' labels for all MCI
subjects.  We then build a CT-NT classifier using as input {\it only} the
image scans at initial visit.\footnote{For a small percentage of the
MCI subjects, we did not obtain the patient's first visit. However,
we did ensure
that the
visit we took as the ``initial visit'' had a CDR of
0.5.}
\section{Results and Discussion} \label{sec:results}
\subsection{Introductory overview}
\label{sec:results-overview} In this section we will perform 1)
classification experiments to evaluate conversion-by-trajectory and
conversion-by-CDR with respect to desideratum 2; 2) additional
experiments to compare the two definitions with respect to
desideratum 3; and lastly, 3) experiments to identify prognostic brain
``biomarker'' regions. 

It is important to mitigate the
potential confounding effect of the subject's age. In our classification experiments, we
mitigated in two ways:

1) For every classifier training, each training sample in one class
was uniquely paired via ``age-matching'' with a training sample in
the other class (with age separation at most one year).

2) For every linear-kernel SVM classifier, we separately
adjusted each feature for age prior to classification using
linear fitting.  We subtracted the extrapolated
line (computed only using ``control'' samples\footnote{For the
AD-Control classifier, these are the samples in the Control class.
For the CT-NT classifier we computed the line using only the NT
samples.}) from the feature's value, for all (training and test)
samples.  As an aside, we note that, given the subsequent linear SVM
operation, the representation power of this linear fitting step is
essentially equivalent to simply treating age as an additional
feature input to the linear SVM classifier.

Finally, prior
to building classifiers, we normalized feature values to the [0,1]
range, which is suitable for the LIBSVM software
\citep{Chang_LIBSVM} we used for training SVMs.

\subsection{Experiments with voxel-based features}
\label{sec:results-voxelbased} The test set (generalization)
accuracy of the voxel-based AD-Control classifier, built using 70
training samples per class, was 0.89 (86 of 88 Controls, and 16 of
27 AD subjects, were correctly classified.) This classifier, with
high specificity for Controls, was then applied to a population of
MCI subjects to determine the CT and NT subgroups. 
\subsubsection{Classification experiments for the MCI population}
\label{sec:results-mci}
\begin{figure*}
\centering \subfigure[By-CDR.] {
    \label{fig:rs}
    \includegraphics[width=1.5 in]{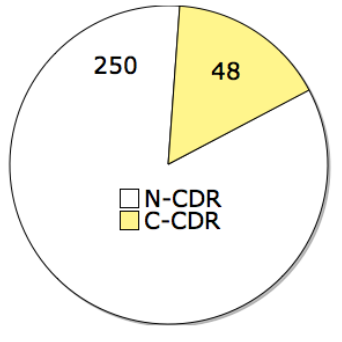}
} \subfigure[By-trajectory.] {
    \label{fig:nc}
    \includegraphics[width=1.5 in]{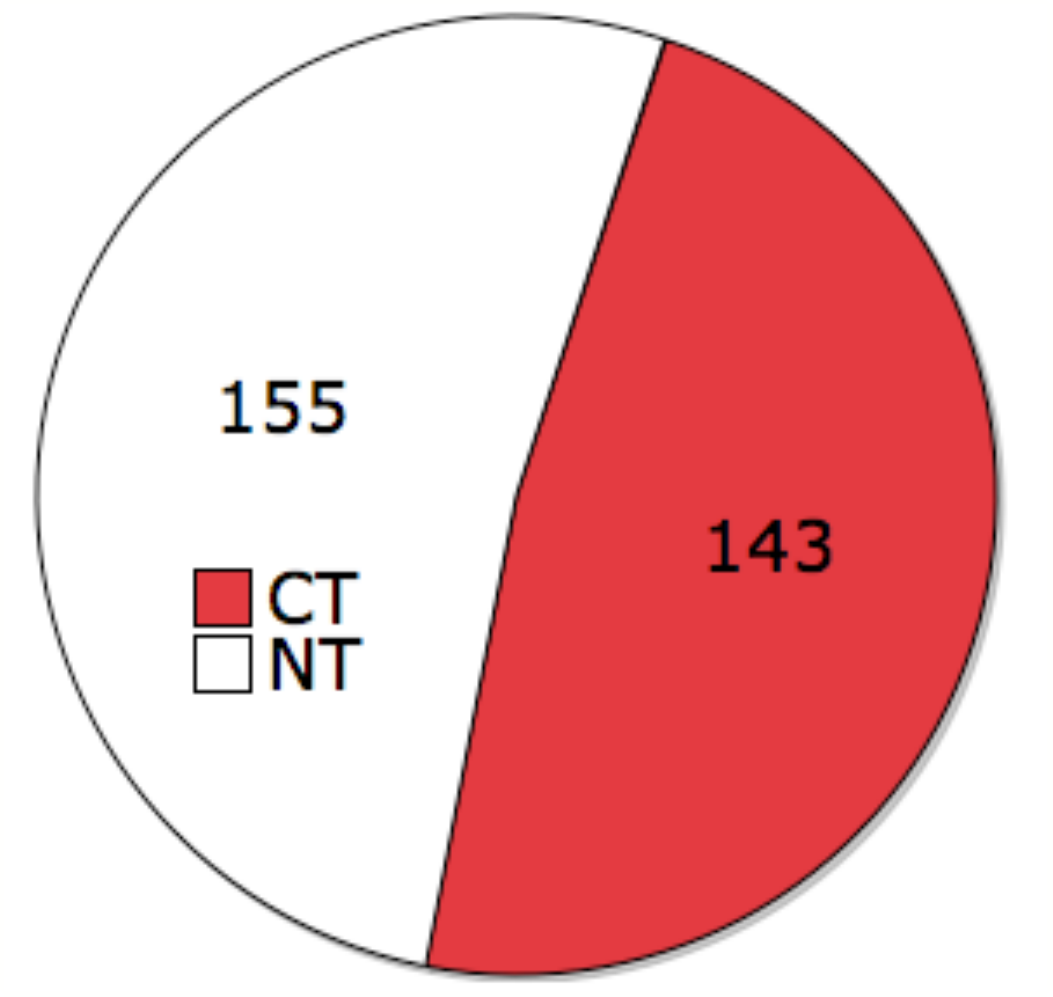}
} \subfigure[Overlap.] {
    \label{fig:four}
    \includegraphics[width=1.5 in]{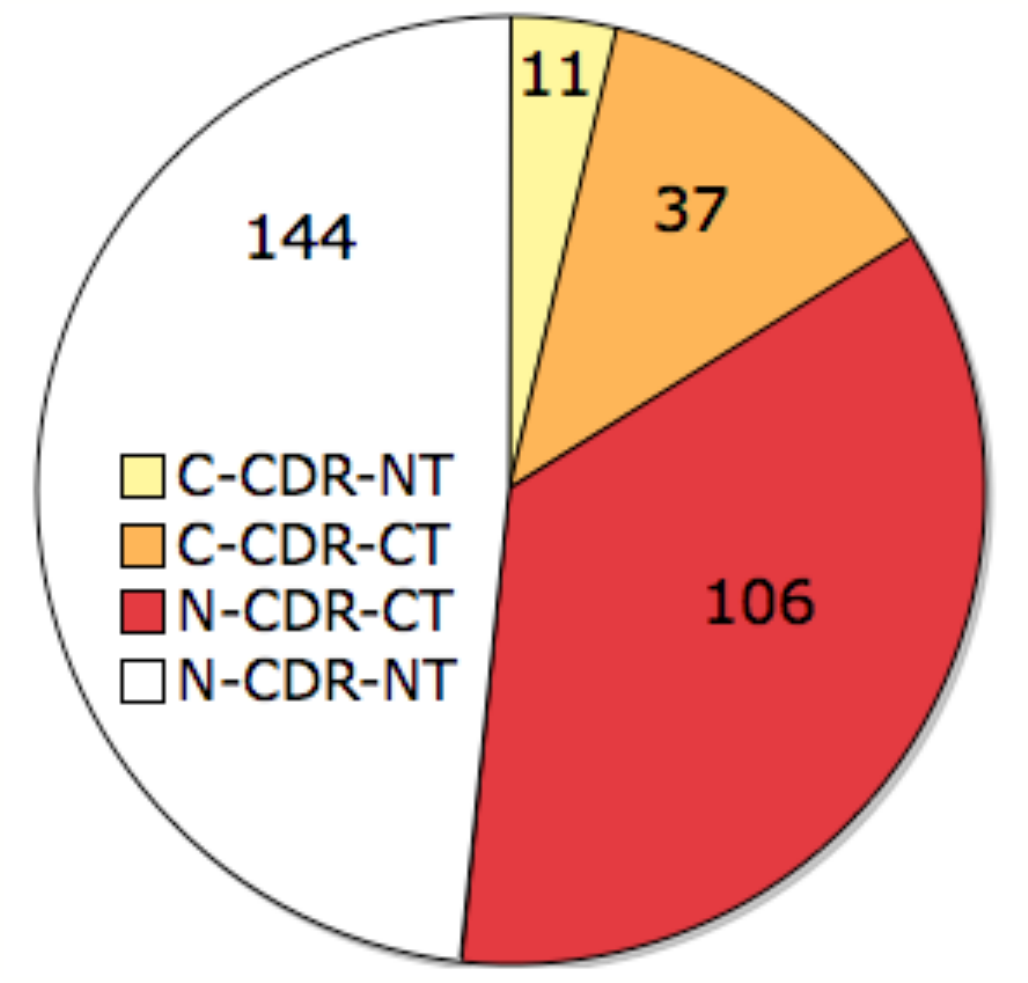}
}\caption{A population of 298 MCI subjects in ADNI
is shown here, broken up according to the two criteria
discussed in Sec. \ref{sec:results-mci}: (a) by-CDR criterion, (b)
by-trajectory criterion; with overlap shown in (c).}
\end{figure*}

Fig. \ref{fig:rs} shows the sizes of the converter-by-CDR (C-CDR)
and nonconverter-by-CDR (N-CDR) groups within the ADNI MCI
cohort for a typical experiment in our work. Fig. \ref{fig:nc}
shows the same population broken up as converters-by-trajectory (CT)
and nonconverters-by-trajectory (NT). Superimposing the two charts,
Fig. \ref{fig:four} illustrates their overlap, where converters by
both definitions are accordingly indicated by orange. Since
converters-by-CDR are relatively scarce, we used a large majority of
them ($80\%$, i.e. 39 individuals among the 48) for the by-CDR
classifier's training set, with the rest ($20\%$) put into the test
set. We reiterate that a general disadvantage of the by-CDR approach
is its scarcity of converter examples -- by contrast, a more
balanced number of examples is available for by-trajectory
training (at least $100$, rather than $39$, training samples
per class, as in Fig. \ref{fig:nc}). Note also that if we were to use a
``probable AD", rather than a by-CDR converter definition, there
would be even {\it fewer} converter examples.

A fair performance comparison between by-trajectory and
by-CDR classification requires: 1) using the same per-class
training set size (i.e. 39) for both by-CDR and by-trajectory
training, {\it and} 2) making the test set sizes the same for both
classifiers.  There are several different ways in which the data can
be partitioned into training and test sets, consistent with these
two conditions: i) we can perform simple random selection on a
class-by-class basis, ensuring only that the two classifiers are
given the same training/test set {\it sizes} (but not the same {\it
sets});\footnote{Note that this means that the training sets for the
converter and nonconverter classes of the conversion-by-CDR
classifier are randomly selected from the yellow and white regions
in Fig. \ref{fig:rs}, respectively, with {\it no} consideration of
trajectory-based (i.e. red/white) labeling illustrated in Fig.
\ref{fig:nc}.} ii) we can make the training sets of the two
classifiers {\it identical} rather than {\it merely same-sized}, as
well as make the test sets identical. This latter approach, though,
will have some bias because, in selecting samples for the
by-trajectory classifier, we will have to make use of knowledge of
the samples' conversion-by-CDR status (and vice versa for the by-CDR
classifier). The first approach, on the other hand, clearly does not
have this bias. As both approaches are valid ways of dealing with
by-CDR data limitations, we will compare generalization accuracies
of by-CDR and by-trajectory classifiers under both these data selection schemes,
referring to these approaches as ``random'' and ``identical'' in the
sequel.

\begin{figure*}
\centering \subfigure[Training/test set selection for by-CDR
classification.] {
    \label{fig:bycdr-biased}
    \includegraphics[width=1.5 in]{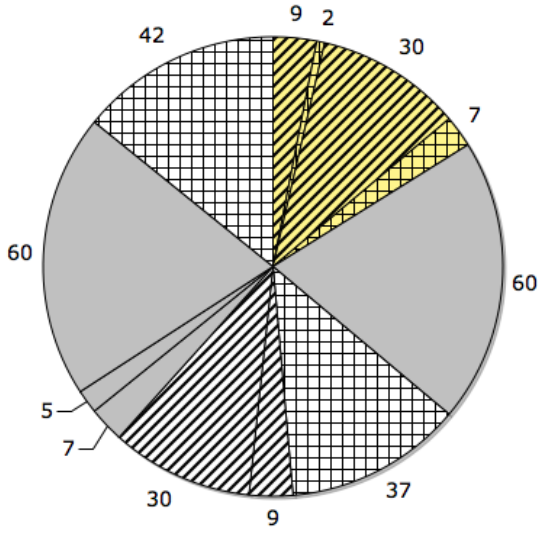}
} \subfigure[Training/test set selection for by-trajectory
classification.] {
    \label{fig:bytraj-biased}
    \includegraphics[width=1.5 in]{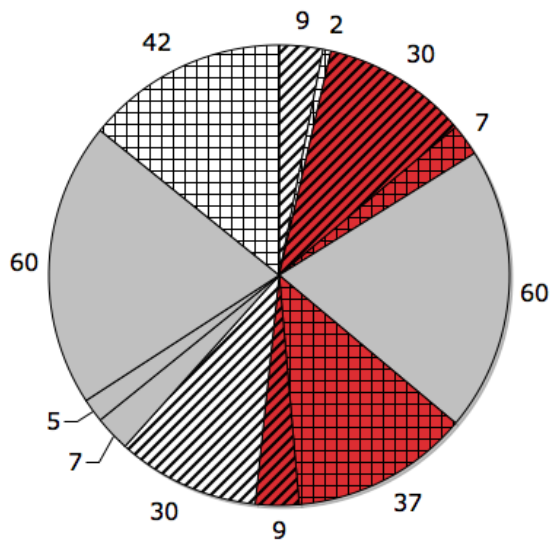} 
} \subfigure[Average test set classification accuracy using all $11,293$ features,
$39\pm 1$ per-class training samples.] {
    \label{table:comparison}
    {\scriptsize
    \begin{tabular}{|l|r|r|r|r|r|r|}
        \hline
        \textbf{Sample} & \textbf{Classifier} & \multicolumn{5}{|c|}{\textbf{Test set}} \\ \cline{3-7}
        \textbf{selection} & & \multicolumn{2}{|c|}{\textbf{Converters}} & \multicolumn{2}{|c|}{\textbf{Nonconverters}} & \textbf{Overall} \\ \cline{3-6}
        & & \textbf{Count} & \textbf{Accuracy} & \textbf{Count} & \textbf{Accuracy} & \textbf{accuracy} \\ \hline
        \textbf{Random} & By-trajectory & $46\pm 1$ & $0.81\pm 0.06$ & $45\pm 1$ & $0.84\pm 0.05$ & \textbf{$0.83\pm 0.05$}  \\ \cline{2-7}
        \textbf{approach} & By-CDR & $9\pm 1$ & $0.79\pm 0.07$ & $82\pm 2$ & $0.47\pm 0.04$ & $0.50\pm 0.04$ \\ \hline
        \textbf{Identical} & By-trajectory & $45\pm 1$ & $0.78\pm 0.04$ & $47\pm 2$ & $0.78\pm 0.04$ & $0.78\pm 0.02$ \\ \cline{2-7}
        \textbf{approach} & By-CDR & $9\pm 1$ & $0.74\pm 0.11$ & $82\pm 2$ & $0.54\pm 0.04$ & $0.56\pm 0.04$  \\ \hline
    \end{tabular}
    }
} \subfigure[By-trajectory. Left: nonconverters; Right: converters] {
    \label{fig:bytraj-nonbiased-f-hist}
    \includegraphics[scale=1.0]{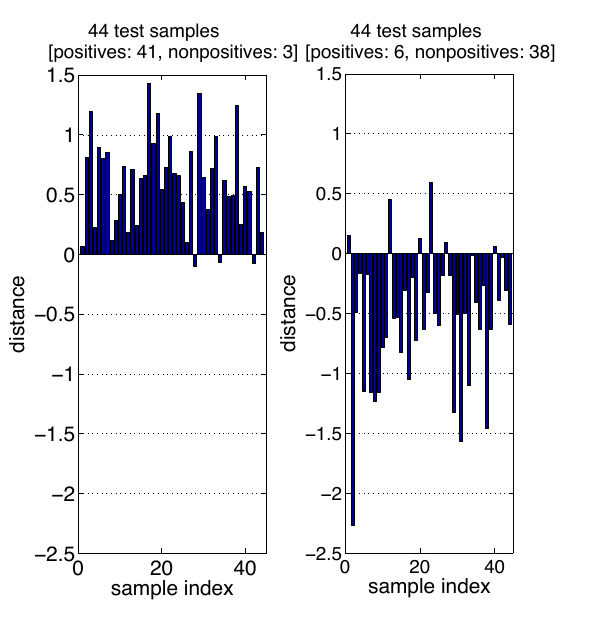}
} \subfigure[By-CDR. Left: nonconverters; Right: converters] {
    \label{fig:bycdr-nonbiased-f-hist}
    \includegraphics[scale=1.0]{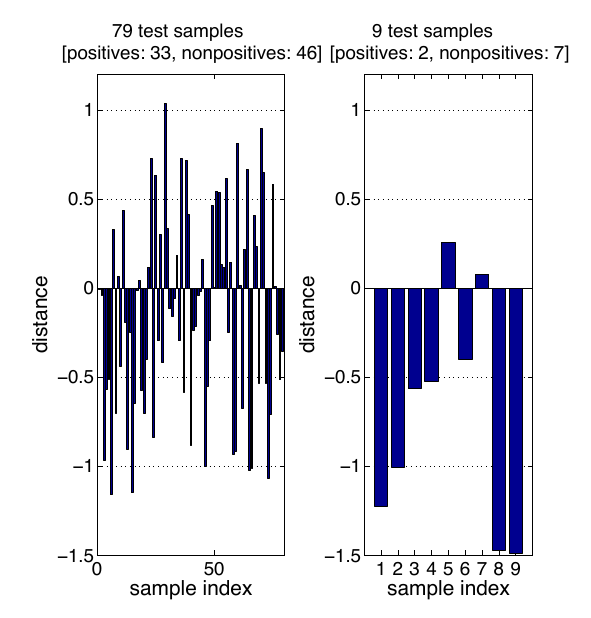}
} \caption{Test set accuracy comparison of by-CDR and by-trajectory
classification.} \label{fig:comparison}
\end{figure*}

Our training/test set selection procedure for the ``identical
approach'' is as follows. For the C-CDR-CT group (Fig.
\ref{fig:four}), randomly select $80\%$ of the group (the yellow
striped group of size 30 in Fig. \ref{fig:bycdr-biased}) such that a
corresponding group within N-CDR (white portion in Fig.
\ref{fig:rs}) can be found that is both {\it NT} and satisfies
age-matching. This corresponding group is illustrated in Fig.
\ref{fig:bycdr-biased} as the white striped group (of size 30),
placed opposite from the yellow striped area it is paired (matched)
with. Likewise for the C-CDR-NT group (Fig. \ref{fig:four}),
randomly select $80\%$ of the group (the yellow striped group of
size 9 in Fig. \ref{fig:bycdr-biased}) such that a corresponding
group within N-CDR can be found that is both {\it CT} and satisfies
age-matching. This corresponding group is illustrated in Fig.
\ref{fig:bycdr-biased} as the white striped group (of size 9).
Notice by comparing this figure to Fig. \ref{fig:four} that the two
white striped areas are separated by the CT-NT border. We take the
training set -- shared by the by-CDR and by-trajectory classifiers
-- to be precisely the union of these four striped
areas.\footnote{For the by-CDR classifier, the class membership of
any of these four subsets of the training set is illustrated by the
color being yellow or white in Fig. \ref{fig:bycdr-biased}.
Likewise, for the by-trajectory classifier, class membership is
illustrated by red or white color in Fig.
\ref{fig:bytraj-biased}.} Subsequently, we take the test set --
shared by the two classifiers -- to be the subjects who are neither
in 1) the training set (striped areas) nor in 2) the special set of
subjects shown in solid gray in Fig. \ref{fig:bycdr-biased} (also
shown identically in Fig. \ref{fig:bytraj-biased}).\footnote{We
exclude this ``special set'' (in gray) from the {\it test set} so
that {\it all} our experiments under the ``identical approach'' can
have a shared, fixed test set (for fair comparison with each other),
including, crucially, an experiment that will include
this  ``special set'' of samples in the {\it training set}.} That
is, the test set is the tiled areas in Fig. \ref{fig:bycdr-biased}
(or, identically, in Fig. \ref{fig:bytraj-biased}).

Note above that {\it some} random selection is being employed in
choosing the training/test sets even in the ``identical approach''
(whereas, in the ``random approach'' the selection is completely
random). Thus, for both approaches, the accuracy of performance
comparison will benefit from averaging accuracy results over
multiple training/test split ``trials''. Results averaged across 10
trials are given in Fig. \ref{table:comparison} for a linear-kernel
SVM\footnote{For generating all classification results herein,
including those in Fig. \ref{table:comparison}, we used SVM
classifiers that were built by employing the common approach of
bootstrap-based validation for selecting the classifier's (trial's)
hyperparameter values \citep{Aksu_TNN}.}; $\mu\pm \sigma$ notation
is used to indicate the mean $\mu$ and standard deviation $\sigma$
of quantities across the trials.\footnote{$\mu$ and $\sigma$ of
integer quantities (e.g. sample counts) are shown rounded up. } 
Note that
by-trajectory's generalization
performance is 
as high as 0.83, whereas by-CDR's generalization performance
is very poor -- {\it as poor as random guessing}
(see 0.5 and 0.56 table values) -- due mainly to poor performance on
{\it nonconverters}-by-CDR.
Fig.
\ref{fig:bytraj-nonbiased-f-hist} and Fig.
\ref{fig:bycdr-nonbiased-f-hist} show by-trajectory and by-CDR
results, respectively, for one of the 10 trials (for the ``random
approach''), with each bar indicating distance to the classification
boundary\footnote{Positive/negative distance means
nonconverter/converter side of the boundary, respectively.} for an
MCI subject in a test population of size 88 and
nonconverters/converters shown in left/right figures, respectively.
Among the 88 subjects, by-trajectory correctly classified 79 whereas
by-CDR correctly classified only 40.

Recently, similarly poor
by-CDR classification performance was also
reported in \citep{Davatzikos_2010}, where it was found
that the majority of
(by-CDR) nonconverters  ``had sharply positive SPARE-AD
scores indicating significant atrophy similar to AD patients''.
Since the SPARE-AD score is produced by a classifier that was
trained to discriminate Control and AD patients \citep{Fan_COMPARE,
Fan_2008}, this comment and associated results are consistent both with
our conjecture in the Introduction and our above histogram results, which
suggest that there may be a significant number of patients undergoing physiological brain
changes consistent with conversion, yet without clinical
manifestation.

The results above indicate that the
conversion-by-CDR definition's two classes are not
well-discriminated, and thus, clinical usefulness of this definition for
our prognostic Aim is expected to be poor. The much greater generalization accuracy
of the by-trajectory definition (coupled with its inherent
plausibility as a conversion definition) indicates its greater
 utility.

\noindent
{\it Increasing the By-Trajectory Image-Based Feature Resolution:}

In a separate
experiment, we evaluated using one of 27 subsamples (rather than one of 216
subsamples), i.e. essentially a 10-fold increase in the number of
(voxel-based) features, and found that the by-trajectory
generalization accuracy rose to 0.91 in the ``random'' case.  We
then tried building 27 separate by-converter classifiers, one for
{\it every} 1/27th subsample (thus effectively using the whole 3D
image), with majority-based voting used to combine the 27 decisions.
This ensemble scheme again achieved 0.91 accuracy, i.e. there was no
further accuracy benefit beyond that from an $\approx$ 10-fold
increase in the number of voxel features. 

\noindent
{\it Increasing the By-Trajectory Training Set Size}

\begin{figure*}
\centering \subfigure[Training/test set selection for by-trajectory
classification.] {
    \label{fig:bytraj-uncon-biased}
    \includegraphics[width=2 in]{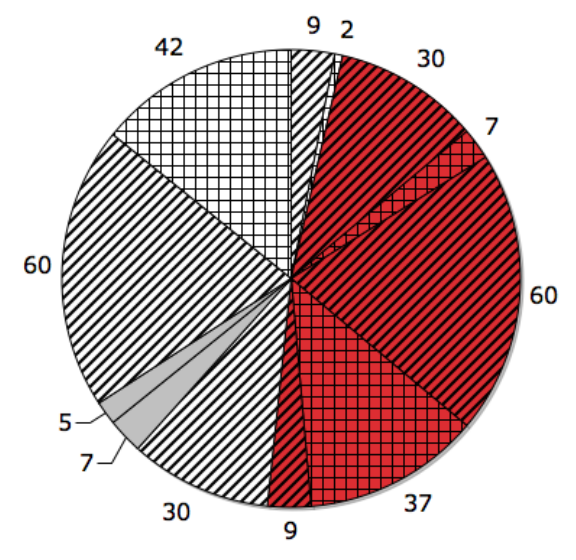} 
} \subfigure[] {
    \label{table:bytraj-uncon}
    {\scriptsize
    \begin{tabular}{|l|r|r|r|r|r|r|}
        \hline
        & \textbf{Sample} & \multicolumn{5}{|c|}{\textbf{Test set}} \\ \cline{3-7}
        & \textbf{selection} & \multicolumn{2}{|c|}{\textbf{Converters}} & \multicolumn{2}{|c|}{\textbf{Nonconverters}} & \textbf{Overall} \\ \cline{3-6}
        & & \textbf{Count} & \textbf{Accuracy} & \textbf{Count} & \textbf{Accuracy} & \textbf{accuracy} \\ \hline
        \textbf{By-trajectory} & \textbf{Random} & $45\pm 1$ & $0.77\pm 0.01$ & $47\pm 2$ & $0.86\pm 0.02$ & $0.82\pm 0.01$  \\ \cline{2-7}
        & \textbf{Identical} & $45\pm 1$ & $0.85\pm 0.03$ & $47\pm 2$ & $0.83\pm 0.05$ & \textbf{$0.84\pm 0.02$} \\ \hline
    \end{tabular}
    }
} \caption{By-trajectory average test set classification accuracy for the
larger training set size ($98\pm 1$ per class) and $11,293$
features.}
\end{figure*}

Note that the converter-by-CDR sample scarcity and class-balancing
(via age-matching) in the experiments above had the effect of
artificially limiting the {\it by-trajectory} classifier training
set size. Next we investigated how much the generalization
accuracy of by-trajectory classification improves when this
limitation is removed. The tiled areas in Figures
\ref{fig:bytraj-uncon-biased} and \ref{fig:bytraj-biased} are
identical, illustrating that in this new experiment (Fig.
\ref{fig:bytraj-uncon-biased}) we used the same test set as
previously, for fairness of comparison. However, as
indicated by differences in the total striped area between these two
charts, we now make the training set much larger than previously.
Specifically, for the ``identical'' case, we used the
previous 10 trials
but simply augmented a trial's
training set with the two large, previously-excluded gray
sets\footnote{Shown in Fig. \ref{fig:bytraj-biased}, with size 60.
Note that this size can vary from trial to trial.}, as these two sets
do age-match each other. The results, averaged across the 10 trials,
are given in Fig. \ref{table:bytraj-uncon}. Notice in this figure
the now larger per-class training size (on average $\approx 100$
rather than $39$), and that the random approach uses this size as
well.
The by-trajectory results
in Fig. \ref{table:bytraj-uncon} indicate that accuracy improved from
$0.78\pm 0.02$ (Fig. \ref{table:comparison}) to $0.84\pm 0.02$ for
the ``identical approach'', and modestly worsened for the
``random approach'' (from $0.83\pm 0.05$ to $0.82\pm 0.01$).

\noindent
{\it Evaluating a Third Conversion Definition}

We now consider a third definition of conversion that combines the
first two definitions as follows.  Let ``converters'' consist of
individuals who converted either by-trajectory {\it or} by-CDR
(non-white areas in Fig. \ref{fig:four}), with the ``nonconverter''
class consisting of the remaining MCI individuals (white area). The
point of view of this new definition, ``conversion-by-union'', is to
be more inclusive in defining an MCI subpopulation at risk, which
may benefit from early treatment or diagnostic testing. While from
that perspective the new definition is reasonable, the fact that
grouping individuals by CDR has a role in this definition may be its
disadvantage, considering that by-CDR classification was previously
shown to perform not much better than random guessing. Results,
averaged across the same 10 trials used in Fig.
\ref{fig:comparison}, are given in Table \ref{table:byunion} and
indicate that conversion-by-union generalizes somewhat worse than
conversion-by-trajectory.\footnote{Note that ``by-union'' is, by
definition, an instance of the ``identical approach'', as stated in
Table \ref{table:byunion}. To ensure fairness of
comparison with the by-trajectory definition of conversion, our test
sets, and training set sizes, in these two cases were identical. In
fact, we chose the by-union training set to be as similar to
by-trajectory's, in every trial, as possible. Referring
to Fig. \ref{fig:bytraj-biased} (which represents a trial example),
the by-union training set was chosen to include 1) the two large
striped groups (red and white); 2) the small ``special'' gray group
(of size seven in this trial example) and its age-matched
counterpart within small white-striped group, and; 3) a subset of
the second small ``special'' gray group (two of five individuals in
this trial example) and its age-matched counterpart within the small
red-striped group.}
\begin{table}
\centerline{
    {\scriptsize
    \begin{tabular}{|l|r|r|r|r|r|r|}
        \hline
        & \textbf{Sample} & \multicolumn{5}{|c|}{\textbf{Test set}} \\ \cline{3-7}
        & \textbf{selection} & \multicolumn{2}{|c|}{\textbf{Converters}} & \multicolumn{2}{|c|}{\textbf{Nonconverters}} & \textbf{Overall} \\ \cline{3-6}
        & & \textbf{Count} & \textbf{Accuracy} & \textbf{Count} & \textbf{Accuracy} & \textbf{accuracy} \\ \hline
        \textbf{By-union} & \textbf{Identical} & $48\pm 1$ & $0.78\pm 0.05$ & $44\pm 2$ & $0.75\pm 0.06$ & $0.77\pm 0.03$  \\ \hline 
    \end{tabular}
    }
} \caption{Average test set accuracy of by-union classification for $39\pm 1$
per-class training samples and $11,293$ features.}
\label{table:byunion}
\end{table}
\subsubsection{Validation on Known AD Conversion Biomarkers}
\label{sec:results-validation} To validate
the proposed conversion definitions with respect to desideratum
3, we performed correlation tests on the MCI
population between the binary class variable $C
\in \{0=no\hspace{0.05in}conversion,\hspace{0.05in}1= conversion\}$
and known AD conversion biomarkers consisting of both 1) volume in
reported AD-affected regions (Table 2 in \citep{Schuff_2010}), which
we measured for each individual's {\it final-visit} MRI\footnote{As
discussed in Appendix A, 
we measured {\it normalized}
region volume. Note also that our regions are defined based on the
atlas (Atlas2) we used. The correspondence between the regions in
\citep{Schuff_2010} and our defined regions is given in
Appendix B.  Finally, note that a subject's final visit is not always
the sixth visit.} and 2) the clinical MMSE measure.
The stronger the
correlation, the more accurately the biomarker is predicted from the
class variable and the greater the separation between the biomarker
histograms, conditioned on the two classes.
\begin{figure*}
\centering
\includegraphics[scale=1.0]{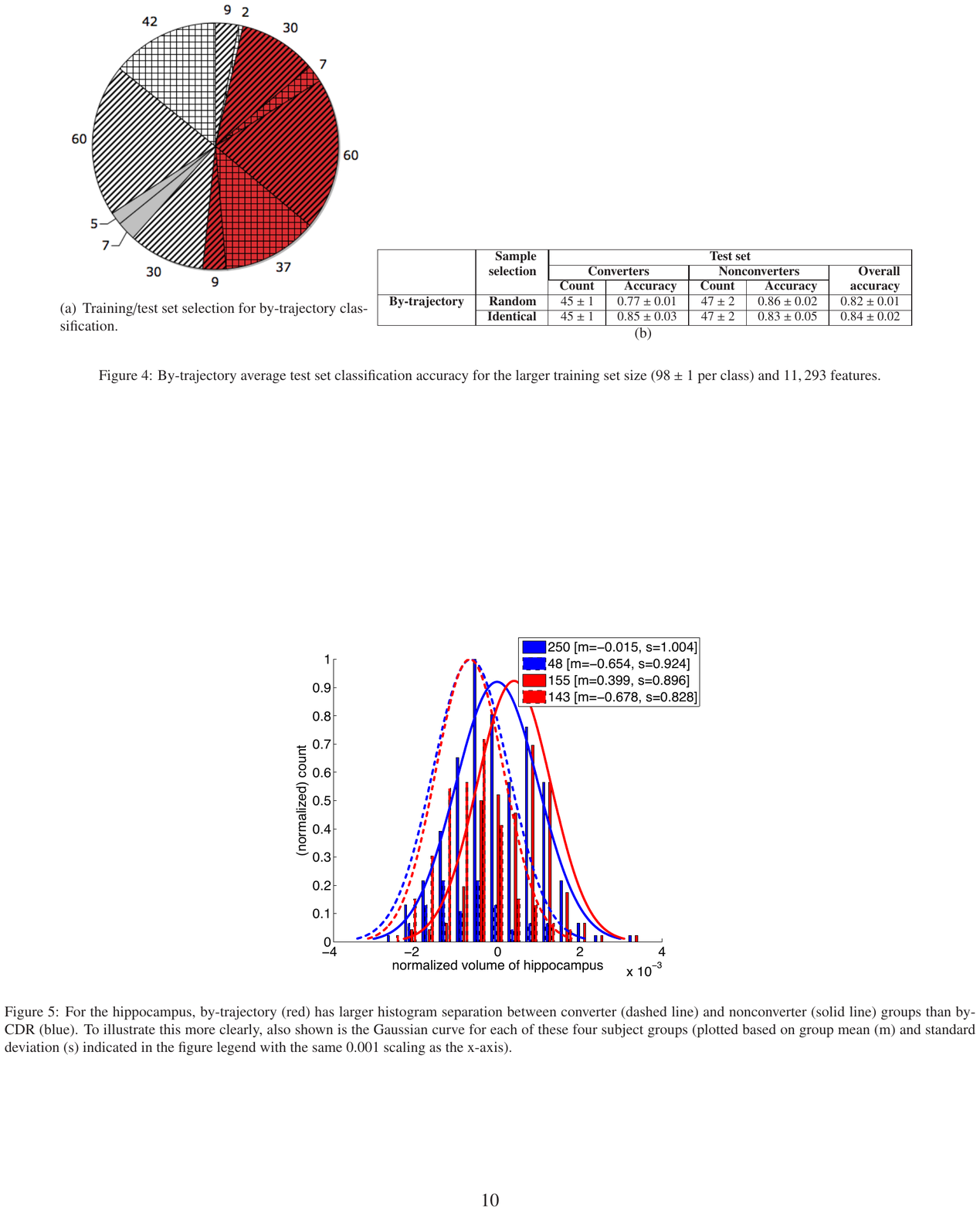}
\caption{For the hippocampus, by-trajectory (red) has larger
histogram separation between converter (dashed line) and
nonconverter (solid line) groups than by-CDR (blue). To illustrate
this more clearly, also shown is the Gaussian curve for each of
these four subject groups (plotted based on group mean (m) and
standard deviation (s) indicated in the figure legend with the same
0.001 scaling as the x-axis).} \label{fig:separation}
\end{figure*}

\begin{figure*}
\centerline{\includegraphics[scale=1.0]{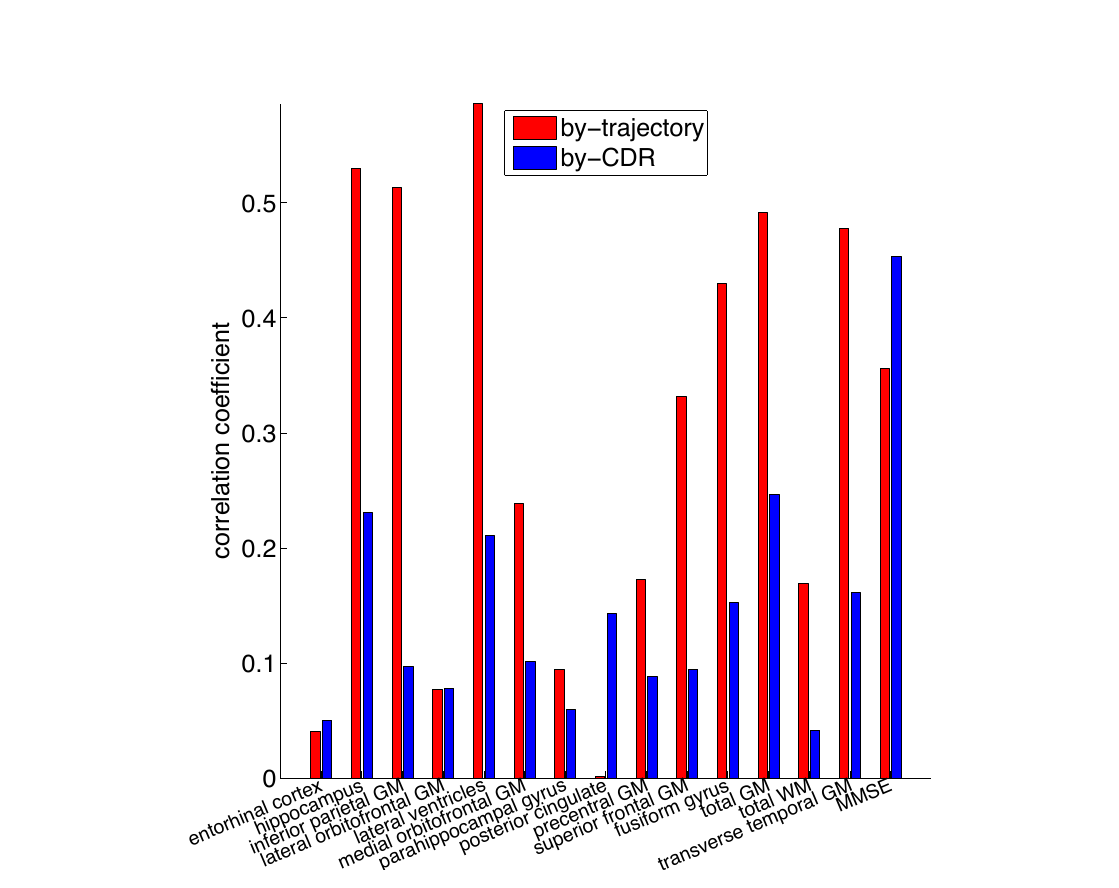}}
\caption{Comparison of correlation coefficients for by-trajectory
and by-CDR.} \label{fig:final-corrected}
\end{figure*}

Before presenting correlation test results, we first
illustrate in Fig. \ref{fig:separation} the increased 
separation of the histograms of hippocampus volume
for the converter and nonconverter groups in the
by-trajectory case, compared with by-CDR.
Next, we
performed comprehensive statistical tests for a number of suggested AD biomarkers. 
The R
statistical computing package was used to perform all tests with
statistical significance set at the 0.05 level. In Table
\ref{table:final-corrected}a (and Fig. \ref{fig:final-corrected}),
the correlation coefficients for by-trajectory and by-CDR are shown
for each biomarker, along with their associated p-values
\citep{Chambers_1993}. Note that for 11 out of 14 brain regions, the
correlation with by-trajectory is greater than the correlation with
by-CDR (in \textbf{bold}), with by-trajectory meeting the significance threshold
in 10 of these 11 regions. Further, for only two of the remaining
four biomarkers - posterior cingulate and the clinical MMSE measure
- does the correlation with by-CDR meet the significance threshold.
Most notably, well-established markers for AD such as the hippocampus,
lateral ventricles, and inferior parietal exhibited strong correlation
with the by-trajectory definition.  

To further assess statistical significance of the {\it comparison}
between by-trajectory and by-CDR correlations, we performed a correlated correlation test
\citep{Meng_1992}, the appropriate test given that the same MCI
sample population was used in measuring correlations for both by-CDR and
by-trajectory.  This test (Table \ref{table:final-corrected}b)
reveals that the larger correlation of by-trajectory is
statistically significant at the 0.05 level in nine brain regions
(in \textbf{bold}).
By contrast, conversion-by-CDR
does not achieve a statistically significant advantage for any of
the brain regions, nor with respect to MMSE.
\begin{table}
\centering \subfigure[] {
    {\scriptsize
    \begin{tabular}{|l|r|r|r|r|}
        \hline
        & \multicolumn{2}{|c|}{\textbf{By-trajectory}} & \multicolumn{2}{|c|}{\textbf{By-CDR}}  \\ \cline{2-5}
        & \textbf{Correlation} & \textbf{P-value} & \textbf{Correlation} & \textbf{P-value}  \\
        \textbf{Biomarker} & \textbf{coefficient} &  & \textbf{coefficient} &   \\ \hline
        Entorhinal cortex  & 0.041 & 0.486  & 0.050 & 0.385 \\ \hline
        Fusiform gyrus &\textbf{0.430}   & 8.20E-15 & \textbf{0.152}   & 0.008 \\ \hline
        Hippocampus & \textbf{0.530}   & 2.00E-16    & \textbf{0.231} &5.67E-05 \\ \hline
        Inferior parietal GM    & \textbf{0.513}   & 2.00E-16    &\textbf{0.097}  & 0.094 \\ \hline
        Lateral orbitofrontal GM    & 0.077  &0.185   & 0.078   & 0.178 \\ \hline
        Lateral ventricles  & \textbf{0.586}   &2.00E-16 & \textbf{0.211} & 0.000246 \\ \hline
        Medial orbitofrontal GM & \textbf{0.239}& 3.01E-05    & \textbf{0.101}   & 0.081 \\ \hline
        Parahippocampal gyrus   &\textbf{0.095}  & 0.102 & \textbf{0.060} & 0.307 \\ \hline
        Posterior cingulate &0.001 & 0.981   & 0.144   & 0.013 \\ \hline
        Precentral GM   &\textbf{0.173}   & 0.003 & \textbf{0.088}   & 0.128 \\ \hline
        Superior frontal GM &\textbf{0.331}   & 4.50E-09 & \textbf{0.094}  & 0.104 \\ \hline
        Superior temporal GM &\textbf{0.478} & 2.00E-16 & \textbf{0.161} & 0.005 \\ \hline
        Total GM    &\textbf{0.491}    & 2.00E-16    & \textbf{0.246}   & 1.70E-05 \\ \hline
        Total WM    &\textbf{0.170}    & 0.003 & \textbf{0.041}  & 0.477 \\ \hline
        MMSE & 0.356    & 2.57E-10   & 0.453 & 2.00E-16 \\ \hline
    \end{tabular}
    }
    } \subfigure[] {
    {\scriptsize
    \begin{tabular}{|r|}
        \hline
        \textbf{P-value}   \\
        \\
        \\ \hline
        0.889 \\ \hline
        \textbf{2.17E-06} \\ \hline
        \textbf{2.42E-10} \\ \hline
        0.985 \\ \hline
        \textbf{1.34E-09} \\ \hline
        \textbf{0.048} \\ \hline
        0.616 \\ \hline
        \textbf{0.045} \\ \hline
        0.230 \\ \hline
        \textbf{0.001} \\ \hline
        \textbf{3.10E-05} \\ \hline
        \textbf{0.0001} \\ \hline
        0.069 \\ \hline
        \textbf{1.33E-06} \\ \hline
        0.118 \\ \hline
    \end{tabular}
    }
} \caption{Correlation coefficients and associated p-values: (a) 
Correlation test results; (b) Correlated correlation test results for
each of the regions in (a). Statistically significant results are shown
in bold.}
\label{table:final-corrected}
\end{table}
To mitigate the confounding effect of age in the AD biomarker
measurements, we linearly adjusted them for age prior to performing
the experiments described above. We also repeated these experiments
without age adjustment. In this case, the number of regions for
which by-trajectory's correlation exceeded by-CDR's rose from 11 to
12 (with the addition of the entorhinal cortex), and the number of
regions for which the larger correlation of by-trajectory was
statistically significant rose from 9 to 10 (with the addition of
the medial orbitofrontal GM).
\subsubsection{Identification of prognostic brain ``biomarker'' regions}
\label{sec:results-MFE}
\begin{table}
\centerline {
    {\scriptsize
    \begin{tabular}{|l|l|l|}
        \hline
        \textbf{AD-Control classifier only} & \textbf{intersection} & \textbf{CT-NT classifier only}  \\ \hline
        Amygdala left & Hippocampal formation* left & Superior temporal gyrus* left \\
        Cingulate region right & Hippocampal formation* right & Middle temporal gyrus left \\
        Entorhinal cortex right & Entorhinal cortex* left & Precuneus right \\
        Inferior occipital gyrus right & Inferior temporal gyrus right & Lateral front-orbital gyrus* right \\
        Medial occipitotemporal gyrus left & Lateral occipitotemporal gyrus* right & Insula right \\
        Parahippocampal gyrus left & Parahippocampal gyrus* right & Supramarginal gyrus* left \\
        Temporal lobe WM right & Perirhinal cortex left & Temporal lobe WM left \\
        Temporal pole right & Perirhinal cortex right & Temporal pole left \\
        & Middle temporal gyrus right & Medial front-orbital gyrus* left \\
        & Uncus left & \\ \hline
    \end{tabular}
    }
    } \caption{Brain regions identified as biomarkers using voxel-based features and MFE.}
\label{table:mfe-voxel-t0}
\end{table}

\begin{figure*}
\centering
\subfigure[]{\label{fig:mfe-curve-ad-voxelbased}\includegraphics[scale=0.3]
{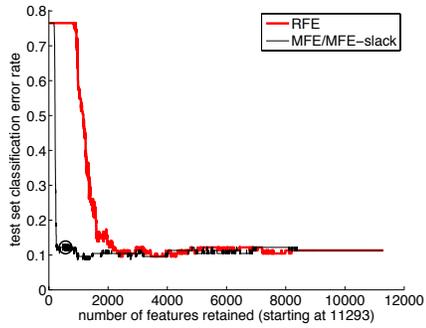}
}
\subfigure[]{\label{fig:mfe-curve-mci-voxelbased}\includegraphics[scale=0.3]
{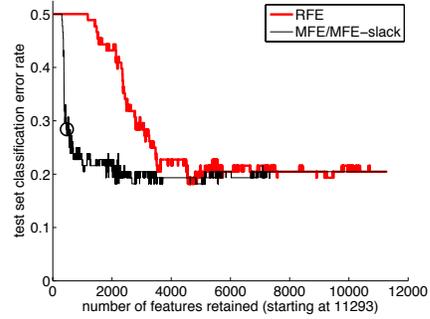}
}
\caption{Test set misclassification rate during the course of
feature elimination, for (a) the AD-Control classifier and (b) the
CT-NT classifier.}
\end{figure*}

In the previous section, we validated conversion definitions using 
established (diagnostic) AD biomarker brain regions (with volumes measured at final visit). 
In this section,
we will {\it identify} key {\it prognostic} biomarker
brain regions via supervised feature selection, aiming first to
identify the ``essential'' subset of voxel features,
i.e. the voxels (at initial visit) necessary for our classifier to
well-discriminate the CT and NT classes. The brain regions
(consistent with a registered brain atlas) within which these select
voxels principally reside then identify our prognostic brain biomarker regions.
Similarly, we will identify {\it diagnostic} regions, critical for discriminating
between AD and Control subjects (using our AD-Control
classifier). In both cases, the accuracy of the selected brain
region biomarkers rests heavily on the accuracy of the supervised
feature selection algorithm we employ. In Figures
\ref{fig:mfe-curve-ad-voxelbased} and
\ref{fig:mfe-curve-mci-voxelbased}, we compare MFE and RFE feature
elimination (i.e. feature selection via feature elimination) for both
Control-AD classification and for CT-NT classification (for one representative,
example trial). The curves show test set
accuracy as a function of the number of retained features (which is
reduced going from right to left). Note that
the ``MFE/MFE-slack'' hybrid method \citep{Aksu_TNN}) outperforms
RFE for both brain classification tasks, achieving lower test set
error rates, and with much fewer retained features. The
circle, determined without use of the test set based on the rule in \citep{Aksu_TNN},
marks the point at which we stopped eliminating features by MFE,
thus determining the (trial's) retained voxel set.  This MFE-RFE comparison (and
the previous comparison in \citep{Aksu_TNN}) supports
our use of MFE to determine brain biomarkers.

\begin{figure*}
\centering \subfigure[] { \label{fig:frac-ad}
    \includegraphics[scale=1.0]{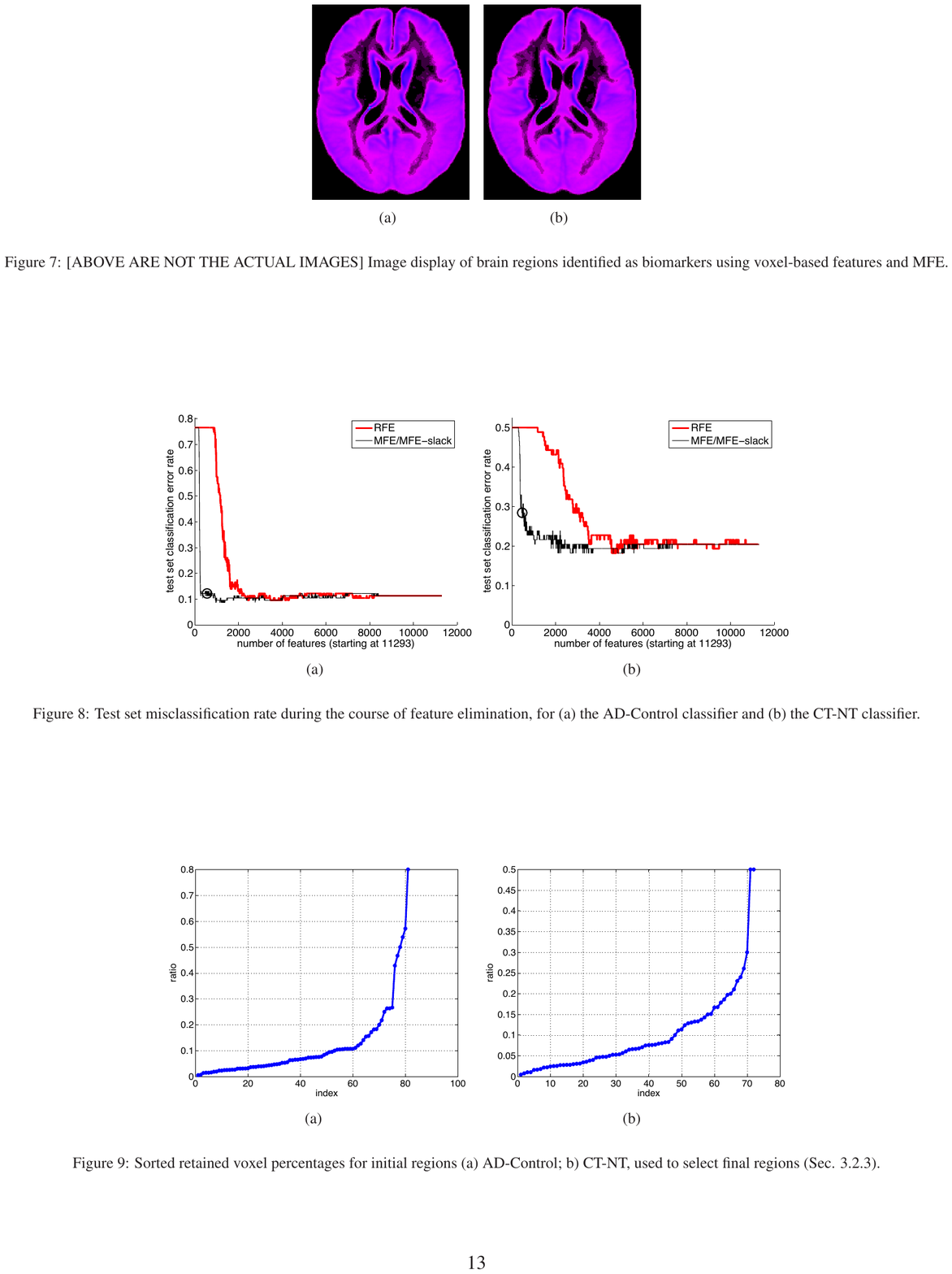}
    } \subfigure[] { \label{fig:frac-ct}
    \includegraphics[scale=1.0]{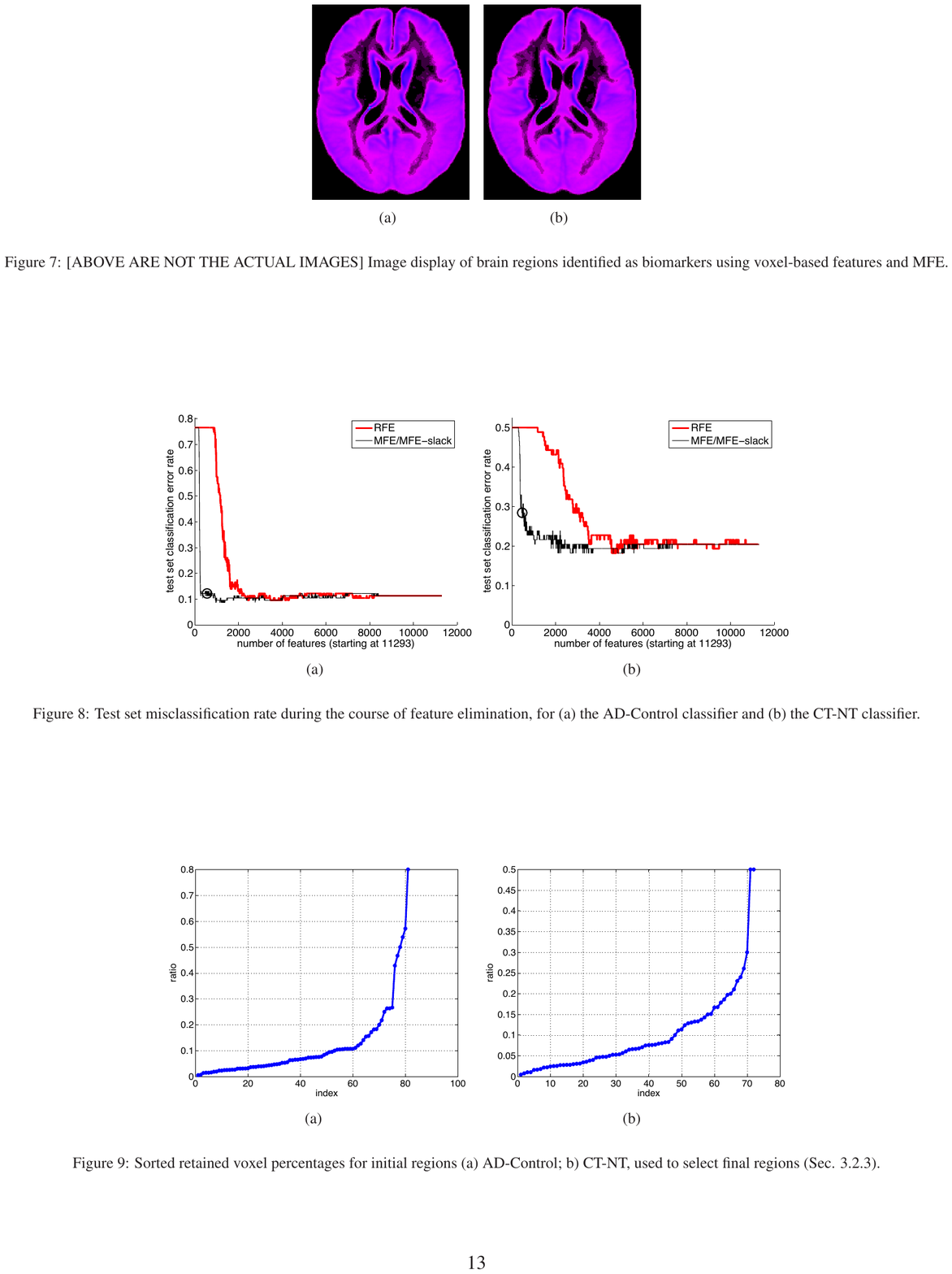}
} \caption{Sorted retained voxel percentages for initial regions (a)
AD-Control; b) CT-NT, used to select final regions (Sec.
\ref{sec:results-MFE}).} \label{fig:frac}
\end{figure*}

To relate the retained voxel set to anatomic regions in the brain, 
we
overlaid the retained voxel set onto a
registered atlas space.
For CT-NT classification, to improve robustness, the final voxel set was formed from 
the union of the retained voxel sets from each of ten feature elimination trials (each using a 
different, randomly selected training sample subset).
For
AD-Control classification, the final
voxel set came from a single trial (the only trial,
from which the 10 CT-NT trials stemmed). 
For
each of these two cases, overlaying the final voxel set onto
the co-registered atlas (Atlas2, defined in 
Appendix A)
yielded between 70-80 anatomic regions.  For data interpretation purposes, we
then identified a subset of (biomarker) regions using the following procedure. 
First, for each brain region, we measured the 
percentage
of the region's voxels that are retained, {\it sorted}
these percentages, and then plotted them. As shown in Fig.
\ref{fig:frac-ad}, the resulting curve for the AD-Control case has a
distinct knee, which we thus used as a threshold (0.125) to select
the final, retained (diagnostic) regions for AD-Control.
We used the same threshold for the CT-NT
curve, shown in Fig. \ref{fig:frac-ct}.  This choice of threshold
yields a reasonable number of regions -- 19 for the CT-NT (prognostic) case
and 21 for the AD-Control (diagnostic) case. 

The resulting sets of identified prognostic and diagnostic biomarkers 
are given in
Table \ref{table:mfe-voxel-t0}, along with their intersection. 
The diagnostic markers in the table include the majority of the
known brain regions in the medial temporal lobe involved in AD pathology.
For example, hippocampus atrophy and lateral ventricle enlargement,
particularly in its anterior aspects of the temporal horn, are 
considered the most prominent diagnostic markers for AD.  Entorhinal
cortical regions, including the perirhinal cortex, are presumably
the earliest sites of degeneration \citep{Braak_1997}.  Thus, independent
identification by our AD-control classifier of known AD diagnostic
biomarkers establishes a reasonable basis for applying the same approach
to identify prognostic biomarkers.  The brain regions listed as CT-NT 
prognostic markers include most known AD diagnostic markers (including 8 of the 12
regions from \citep{Schuff_2010} (marked by *), 4 of which are also diagnostic
markers), indicating
that some AD-linked pathological changes in these brain regions already
occurred and remained active in a subset of MCI subjects who likely progress
to AD rapidly.  
Conversely, the brain areas appearing only on the prognostic
marker list are likely the most active areas of degeneration during this stage
of progression to dementia.  These structures tend to be the brain regions
further away from the entorhinal cortex onto the parietal (Supramarginal gyrus,
Precuneus) and temporal cortex (Superior temporal gyrus and Middle temporal
gyrus) regions.
All the brain structures listed in the table are known to be involved in 
AD \citep{Braak_1997,Chan_2003,Frisoni_2009}.  Thus, the markers in Table 3 suggest an interesting anatomic
pattern of trajectory for MCI conversion to AD which conforms with the Brak and
Brak hypothesis and previous imaging findings \citep{Chan_2003,Frisoni_2009}.
Moreover, the CT-NT regions uniquely found by our MFE-based
procedure in Table \ref{table:mfe-voxel-t0} may be viewed as
``putative'' prognostic markers, and may warrant further investigation.

Finally, we note that we have used a particular criterion (percentage of a region's voxels that are
retained) to identify biomarker regions, starting from MFE-retained voxels.  While our 
identified regions are plausible, it is possible that other (equally reasonable) criteria may produce
different biomarker region results.  Thus, the biomarkers we identify should be viewed as anecdotal,
identifying regions that figure prominently in our classifier's decisionmaking and also potentially assisting
researchers in forming hypotheses about MCI-to-AD disease progression. However, we do not view the identified
regions as definitive.

\subsubsection{Comparison with an SPM-based biomarker identification approach}
\label{sec:results-SPM} In the previous section, we used MFE to
identify voxels as biomarkers for the CT and NT classes. Here,
using the same CT-NT training and test populations, we will
alternatively identify voxel-based biomarkers using statistical testing
with SPM5 (see: \citep{SPM_website}). Subsequently we will present a
classifier generalization accuracy comparison (where accuracy is
again measured on the previous section's CT-NT ({\it test}) population)
for these two biomarker detection methods. We determined SPM
biomarkers as follows. The CT-NT training set population, being
age-matched, is readily suitable for a paired t-test, an appropriate
statistical test for determining SPM-identified biomarkers, i.e.
voxels that discriminate between the CT and NT groups. In contrast
with MFE's use of {\it only one} (out of 216) RAVENS subsamples
(taken jointly from the GM, WM, ventricle maps), we performed
t-tests on {\it whole} RAVENS maps (without subsampling), which
makes the SPM-MFE comparison favorably biased towards SPM.
More specifically our steps were as follows. First, for the GM and
WM maps separately, we found using SPM that a large portion of each
of these two tissues was statistically significant at the 0.05 level
when correction for multiple comparisons was not applied. Next, we
used SPM's FDR-based correction for multiple comparisons -- based on
an SPM FDR cluster size of 5 voxels we found that the spatial extent
of the statistically significant regions, at each of the levels
0.05, 0.01, and 0.005, was 
approximately a subset 
of the above-mentioned spatial support found in the uncorrected case.
Given that the number of significant voxels in any of these SPM
experiments is, again due to no subsampling, much larger than the 11,293
voxels started from in the MFE case,
we simply 1) chose as our SPM result the result for 0.01
(FDR-corrected),
2) took from among those significant voxels the {\it most} significant
11,434 voxels in order to be able to compare MFE and SPM for
$\approx$ the same number of voxels (biomarkers).
To obtain the generalization accuracy for this
SPM-identified biomarker voxel set, using the same training/test set
as in the MFE experiment, we trained an SVM classifier and measured
its generalization accuracy, which was found to be 0.76. This
accuracy
is somewhat lower than the 0.8 accuracy of the
previous section's CT-NT SVM classifier. Recalling that this
comparison is actually favorably biased towards SPM, and further
noticing the fact that MFE was able to maintain the 0.8 accuracy all
the way down to 2000 features (cf. Fig.
\ref{fig:mfe-curve-mci-voxelbased}), this experimental comparison
provides another validation (beyond the comparison with RFE given earlier)
for MFE-based feature/biomarker selection, applied to brain images.
\section{Conclusions} \label{sec:conc}
We have presented an automated prognosticator of
MCI-to-AD conversion based on brain morphometry derived from high resolution
ADNI MR images.  The primary novel contributions of our work are:
i) casting MCI prognostication as a novel machine learning problem lying
somewhere between supervised and unsupervised learning; ii) our proposal
of a conversion definition which, unlike previous methods, exploits both rich
phenotypical information in neuroimages and AD and control examples; iii) correlation
testing and classifier accuracy evaluations to validate candidate conversion
definitions; iv) prognostic biomarker discovery based on our conversion definition.
We
demonstrated that our method achieved both better
generalization accuracy and stronger, statistically significant,
correlations with known brain region biomarkers than a predictor
based on the clinical CDR score, the approach used in
several past works. 
The brain structures identified as AD-control diagnostic markers and
MCI conversion prognostic markers well conform with known brain atrophic 
patterns and progression trajectories occurring in AD-afflicted brains.
While the noisy nature of cognitive assessments,
including MMSE, has been acknowledged in past works, in future we
may extend our methodology to consider cognitive assessment data,
both potentially as additional (baseline) input features and as
additional or alternative prediction targets to our
``conversion-by-trajectory'' labels.  We may also consider
alternative ways to adjust for confounding effects of age, noting that \citep{Schuff_2010}
has characterized the nonlinear dependence of age on brain region
volumes. 
Finally, while we have focused on the
MCI subpopulation here, our system could also potentially be used to
detect, as possible misdiagnoses, subjects diagnosed as ``Control''
who are classified as MCI converters by our system.
\section{Acknowledgement} \label{sec:ad-ack}
Funding supporting this work has been provided in part through NIH R01 AG02771 and
the Pennsylvania Department of Health.  
We thank Dr. Michelle Shaffer for assistance with statistical testing and
Jianli Wang and Zachary Herse for assistance with segmentation.

Data collection and sharing for this project was funded by the
Alzheimer's Disease Neuroimaging Initiative (ADNI) (National
Institutes of Health Grant U01 AG024904). ADNI is funded by the
National Institute on Aging, the National Institute of Biomedical
Imaging and Bioengineering, and through generous contributions from
the following: Abbott, AstraZeneca AB, Bayer Schering Pharma AG,
Bristol-Myers Squibb, Eisai Global Clinical Development, Elan
Corporation, Genentech, GE Healthcare, GlaxoSmithKline,
Innogenetics, Johnson and Johnson, Eli Lilly and Co., Medpace, Inc.,
Merck and Co., Inc., Novartis AG, Pfizer Inc, F. Hoffman-La Roche,
Schering-Plough, Synarc, Inc., as well as non-profit partners the
Alzheimer's Association and Alzheimer's Drug Discovery Foundation,
with participation from the U.S. Food and Drug Administration.
Private sector contributions to ADNI are facilitated by the
Foundation for the National Institutes of Health (www.fnih.org). The
grantee organization is the Northern California Institute for
Research and Education, and the study is coordinated by the
Alzheimer's Disease Cooperative Study at the University of
California, San Diego. ADNI data are disseminated by the Laboratory
for Neuro Imaging at the University of California, Los Angeles. This
research was also supported by NIH grants P30 AG010129, K01
AG030514, and the Dana Foundation.
\appendix
\section{Image processing} \label{sec:image-proc}
The input to
our processing is a $T_{1}$-weighted MR image of the head. First,
using rigid-body registration implemented in FSL's linear image
registration tool FLIRT \citep{FSL-FLIRT}, we coarsely aligned this
(3d) image with the MNI/ICBM
atlas resampled to 1mm isotropic voxel dimensions (aka Atlas1). We
used FSL version 3 (see: \citep{FSL_website}). Next, we removed
non-brain anatomy from the aligned head image, using FSL's brain
extraction tool BET \citep{FSL-BET}. We then segmented the resulting
brain-only 3d image into the following five segments required by
HAMMER\citep{Shen_HAMMER_ITMI}\footnote{We used a 2006
version of HAMMER, downloaded on November 8, 2006.}
: WM, GM, ventricles, (non-ventricle) CSF, (non-anatomy)
background.
Next, using as inputs
an MNI atlas distributed with HAMMER (aka Atlas2\footnote{Since the
region-segmented version of the atlas distributed with HAMMER had
much better segmentation quality than the five-segment version, as a
pre-processing step we used the former to create a replacement for
the latter (by simply taking the union of all the regions). The
resulting five-segment atlas is dubbed `Atlas2'.}) and the ADNI
participant's five-segment image, we performed two different HAMMER
operations, generating: 1) the
three ``volumetric density'' Ravens images: GM, WM, ventricles.  The union of these
three images (following some preprocessing) forms the set of features used
by our classifier; 2) 
the 3d region-segmented image, 
whose region volumes are 
used for by-trajectory statistical validation in
section 3.2.2 and for region biomarker identification in section 
3.2.3.\footnote{We measured normalized region volumes. We normalized
by dividing by the sum of all (98) region volumes. This sum is
essentially intracranial volume minus cerebellum volume, as our
``intracranial region" list includes CSF in addition to brain regions
and excludes the cerebellum.}

Each of these two HAMMER operations performs a different type of atlas-based
nonlinear registration -- the one that generates the RAVENS
tissue maps performs less aggressive warping between the
five-segment image and the atlas in order both to combat noise
inherent in the registration process and to preserve volume on a
tissue-by-tissue basis (so as to properly detect and encode brain
atrophy). Consequently, there is considerable individual variability
in RAVENS images, which we mitigate in a standard way by smoothing with a
Gaussian filter with an FWHM
of 5mm.

The sum of the voxel intensities over the three RAVENS maps is on
the order of $10^6$ and varies across individuals.  Thus, prior to
smoothing, we normalized each individual's Ravens maps, such that
each individual has the same total volume.
However, after the normalization and smoothing (aka ``nsRAVENS''
images), some areas of poor registration, manifesting as
areas with very low voxel values (including zero), will
remain.\footnote{A substantial portion of the very low, nonzero
voxel intensities are in fact introduced by the smoothing itself as
it calculates each new voxel intensity as a weighted average of many
neighboring voxels (thereby switching some voxels from zero
(non-anatomy) to very low intensity values (anatomy), which
essentially slightly grows the anatomy boundaries outward).} For a
population of nsRAVENS images, these areas are considered to be outlier
areas and are thus removed from each image in the population by thresholding
\footnote{We calculated the
threshold solely using the {\it training population} of our
AD-Control classifier, and then applied the thresholding operation
on the entire population of AD, MCI, and Control individuals. In
this way, we were careful to exclude test examples from all phases
of classifier training.}.
The resulting images are the features used by our classification system.

\section{Correspondence Between Atlas-defined Regions and Those Defined in (Schuff et al., 2010)}

\begin{table}[h!]
\centerline{
    {\scriptsize
    \begin{tabular}{|l|l|}
        \hline
        Entorhinal cortex & Entorhinal cortex left/right \\ \hline
        Fusiform gyrus & Lateral occipitotemporal gyrus right/left \\ \hline
        Hippocampus & Hippocampal formation right/left \\ \hline
        Inferior parietal GM & Supramarginal gyrus left/right, Angular gyrus right/left \\ \hline
        Lateral orbitofrontal GM & Lateral front-orbital gyrus right/left \\ \hline
        Lateral ventricles & Lateral ventricle left/right \\ \hline
        Medial orbitofrontal GM & Medial front-orbital gyrus right/left \\ \hline
        Parahippocampal gyrus & Parahippocampal gyrus right/left \\ \hline
        Posterior cingulate & Cingulate region left/right \\ \hline
        Precentral GM & Precentral gyrus right/left \\ \hline
        Superior frontal GM & Superior frontal gyrus left/right \\ \hline
        Superior temporal GM & Superior temporal gyrus right/left \\ \hline
    \end{tabular}
    }
} \caption{Correspondence between the regions in \citep{Schuff_2010}
(left) (except ``Total GM'' and ``Total WM'') and our defined
regions (right).} \label{table:corresp}
\end{table}
\bibliographystyle{elsarticle-harv}

\newpage
\begin{thebibliography}{00}




\bibitem[Aksu et al.(2010)]{Aksu_TNN}
Y. Aksu, D. J. Miller, G. Kesidis, and Q. X .Yang,
``Margin-Maximizing Feature Elimination Methods for Linear and
Nonlinear Kernel-Based Discriminant Functions'', {\it IEEE
Transactions on Neural Networks}, vol. 25, no.10, pp.701-717, 2010.

\bibitem[Braak et al.(1997)]{Braak_1997}
H. Braak and E. Braak,
``Frequency of Stages of Alzheimer-Related Lesions in Different Age
Categories'', {\em Neurobiology of Aging} {\bf 18}(4):351-357, July 8 1997.


\bibitem[Chang et al.(2001)]{Chang_LIBSVM}
C. Chang and C. Lin, ``LIBSVM : a library for support vector
machines,'' software available at
http://www.csie.ntu.edu.tw/$\sim$cjlin/libsvm, 2001.

\bibitem[Chambers et al.(1993)]{Chambers_1993}
J. M. Chambers, ``Linear models'', in: {\em Statistical Models in
S}, J. M. Chambers and T. J. Hastie (Eds.), Chapman \& Hall, New
York, 1993.

\bibitem[Chan et al. (2003)]{Chan_2003}
D. Chan, J.C. Janssen, J.L. Whitwell, H.C. Watt, R. Jenkins, C. Frost,
M.N. Rossor and N.C. Fox, ``Change in rates of cerebral atrophy over time
in early-onset Alzheimer's disease: longitudinal MRI study", {\em Lancet}
{\bf 362}, pp. 1121-1122, 2003.


\bibitem[Chetelat et al.(2002)]{Chetelat_2002}
G. Chetelat, B. Desgranges, V. de la Sayette, F. Viader, F.
Eustache, J-C. Baron, ``Mapping gray matter loss with voxel-based
morphometry in mild cognitive impairment'', {\it NeuroReport}
vol.13. no.15, 28 October 2002.

\bibitem[Chou et al.(2009)]{Chou_SPIE}
Y.-Y. Chou, N. Lepor$\acute{e}$, C. Avedissian, S. K. Madsen, X.
Hua, C. R. Jack Jr., M. W. Weiner, A. W. Toga, P. M. Thompson, and
the Alzheimer's Disease Neuroimaging Initiative, ``Mapping
Ventricular Expansion and its Clinical Correlates in Alzheimer's
Disease and Mild Cognitive Impairment using Multi-Atlas Fluid Image
Alignment'', {\it Proc. SPIE}, vol.7259, 725930, 2009.

\bibitem[Chou et al.(2010)]{Chou_ADNI}
Y.-Y. Chou, N. Lepor$\acute{e}$, P. Saharan, S. K. Madsen, X. Hua,
C. R. Jack, L. M. Shaw, J. Q. Trojanowski, M. W. Weiner, A. W. Toga,
P. M. Thompson, and the Alzheimer's Disease Neuroimaging Initiative,
``Ventricular maps in 804 ADNI subjects: correlations with CSF
biomarkers and clinical decline'', {\it Neurobiology of Aging} 31,
pp.1386-1400, 2010.

\bibitem[Csernansky et al.(2005)]{Csernansky_2005}
J. G. Csernansky, L. Wang, J. Swank, J. P. Miller, M. Gado, D.
McKeel, M. I. Miller, J. C. Morris, ``Preclinical detection of
Alzheimer's disease: hippocampal shape and volume predict dementia
onset in the elderly'', {\it NeuroImage} 25, pp.783-792, 2005.

\bibitem[Davatzikos(1998)]{Davatzikos_1998}
C. Davatzikos, ``Mapping image data to stereotaxic spaces:
Applications to brain mapping,'' {\it Hum. Brain Mapp.}, 6:334-338,
1998.

\bibitem[Davatzikos et al.(2001)]{Davatzikos_2001}
C. Davatzikos, A. Genc, D. Xu, and S. M. Resnick, ``Voxel-Based
Morphometry Using the RAVENS Maps: Methods and Validation Using
Simulated Longitudinal Atrophy,'' {\it NeuroImage} 14, pp.1361-1369,
2001.

\bibitem[Davatzikos et al.(2008)]{Davatzikos_2008}
C. Davatzikos, Y. Fan, X. Wu, D. Shen, S. M. Resnick, ``Detection of
prodromal Alzheimer's disease via pattern classification of magnetic
resonance imaging'', {\it Neurobiology of Aging} 29, pp.514-523,
2008.

\bibitem[Davatzikos et al.(2010)]{Davatzikos_2010}
C. Davatzikos, P. Bhatt, L. M. Shaw, K. N. Batmanghelich, J. Q.
Trojanowski, ``Prediction of MCI to AD conversion, via MRI, CSF
biomarkers, and pattern classification'', {\it Neurobiology of
Aging}, 2010, doi:10.1016/j.neurobiolaging.2010.05.023.

\bibitem[de Leon et al.(2006)]{DeLeon_2006}
M. J. de Leon, S. DeSanti, R. Zinkowski, P. D. Mehta, D. Pratico, S.
Segal, H. Rusinek, J. Li, W. Tsui, L. A. Saint Louis, C. M. Clark,
C. Tarshish, Y. Li, L. Lair, E. Javier, K. Rich, P. Lesbre, L.
Mosconi, B. Reisberg, M. Sadowski, J. F. DeBernadis, D. J. Kerkman,
H. Hampel, L. -O. Wahlund, P. Davies, ``Longitudinal CSF and MRI
biomarkers improve the diagnosis of mild cognitive impairment,''
{\it Neurobiology of Aging} 27, pp.394-401, 2006.

\bibitem[De Meyer et al.(2010)]{DeMeyer_2010}
G. De Meyer, F. Shapiro, H. Vanderstichele, E. Vanmechelen, S.
Engelborghs, P. P. De Deyn, E. Coart, O. Hansson, L. Minthon, H.
Zetterberg, K. Blennow, L. Shaw, J. Q. Trojanowski, for the
Alzheimer's Disease Neuroimaging Initiative, ``Diagnosis-Independent
Alzheimer Disease Biomarker Signature in Cognitively Normal Elderly
People'', {\it Arch. Neurol.}, vol. 67, no. 8, pp.949-956, Aug 2010.

\bibitem[Duda et al.(2001)]{Duda}
R. Duda, P. Hart, and G. Stork, {\em Pattern Classification.} Second
Edition, John Wiley and Sons, New York, 2001.

\bibitem[Fan et al.(2007)]{Fan_COMPARE} Y. Fan, D. Shen, R. C. Gur, R. E. Gur,
C. Davatzikos, ``COMPARE: Classification of Morphological Patterns
Using Adaptive Regional Elements'', {\it IEEE Transactions on
Medical Imaging}, vol. 26, no. 1, pp.93-105, 2007.

\bibitem[Fan et al.(2008)]{Fan_2008}
Y. Fan, N. Batmanghelich, C. M. Clark, C. Davatzikos, ``Spatial
patterns of brain atrophy in MCI patients, identified via
high-dimensional pattern classification, predict subsequent
cognitive decline'', {\it NeuroImage} 39, pp.1731-1743, 2008.

\bibitem[Fennema-Notestine et al.(2009)]{Fennema_2009}
C. Fennema-Notestine, D. J. Hagler Jr., L. K. McEvoy, A. S.
Fleischer, E. H. Wu, D. S. Karow, A. M. Dale, the Alzheimer's
Disease Neuroimaging Initiative, ``Structural MRI biomarkers for
preclinical and mild Alzheimer's disease'', {\it Human Brain
Mapping}, vol.30, issue 10, pp.3238-3253, 2009.

\bibitem[Frisoni et al.(2009)]{Frisoni_2009}
G.B. Frisoni, A. Prestia, P.E. Rasser, M. Bonetti, P.M. Thompson,
``In vivo mapping of incremental cortical atrophy from incipient to overt
Alzheimer's disease", {\em Neurol.} {\bf 256}(6):916-24, Feb 28, 2009.


\bibitem[FSL website()]{FSL_website}
FSL website: FSL (FMRIB Software Library),
http://www.fmrib.ox.ac.uk/fsl

\bibitem[Fung et al.(2004)]{Fung-NLPSVM}
G. Fung, O. L. Mangasarian, ``A feature selection Newton method for
support vector machine classification,'' {\it Computational
Optimization and Applications}, vol. 28, no.2:185-202, July 2004.

\bibitem[Goldszal et al.(1998)]{Goldszal_1998}
A. F. Goldszal, C. Davatzikos, D. L. Pham, M. X. H. Yan, R. N.
Bryan, S. M. Resnick, ``An Image-Processing System for Qualitative
and Quantitative Volumetric Analysis of Brain Images,'' {\it J.
Comput. Assist. Tomogr.}, 22:827-837, 1998.

\bibitem[Guyon et al.(2002)]{Guyon_RFE}
I. Guyon, J. Weston, S. Barnhill, and V. Vapnik, ``Gene selection
for cancer classification using support vector machines,'' {\it
Machine Learning}, 46(1):389-422, 2002.

\bibitem[Guyon et al.(2003)]{Guyon_2003}
I. Guyon and A. Elisseeff, ``An introduction to variable and feature
selection,'' {\it J. Mach. Learn. Res.}, vol. 3, pp.1157-1182, 2003.

\bibitem[Jenkinson et al.(2001)]{FSL-FLIRT}
M. Jenkinson and S.M. Smith, ``A global optimisation method for
robust affine registration of brain images,'' {\it Med. Image
Anal.}, 5(2):143-156, 2001.

\bibitem[Lerch et al.(2005)]{Lerch_2005}
J. P. Lerch, A. C. Evans, ``Cortical thickness analysis examined
through power analysis and a population simulation'', {\it
NeuroImage} 24, pp.163-173, 2005.

\bibitem[Leow et al.(2009)]{Leow_2009}
A. D. Leow, I. Yanovsky, N. Parikshak, X. Hua, S. Lee, A. W. Toga,
C. R. Jack Jr., M. A. Bernstein, P. J. Britson, J. L. Gunter, C. P.
Ward, B. Borowski, L. M. Shaw, J. Q. Trojanowski, A. S. Fleisher, D.
Harvey, J. Kornak, N. Schuff, G. E. Alexander, M. W. Weiner, P. M.
Thompson, Alzheimer's Disease Neuroimaging Initiative. ``Alzheimer's
Disease Neuroimaging Initiative: A One-Year Follow-up Study Using
Tensor-Based Morphometry Correlating Degenerative Rates, Biomarkers
and Cognition,'' {\it Neuroimage} 45, pp.645-55, 2009.

\bibitem[Meng et al.(1992)]{Meng_1992}
X. -L. Meng, R. Rosenthal, D. B. Rubin, ``Comparing Correlated
Correlation Coefficients'', {\it Psychological Bulletin} 111,
pp.172-175, 1992.

\bibitem[Misra et al.(2009)]{Misra_2009} C. Misra, Y. Fan, C. Davatzikos,
``Baseline and longitudinal patterns of brain atrophy in MCI
patients, and their use in prediction of short-term conversion to
AD: Results from ADNI'', {\it NeuroImage} 44, pp.1415-1422, 2009.

\bibitem[Petersen(2004)]{Petersen_2004}
R. C. Petersen, {\em Mild cognitive impairment: aging to Alzheimer's
Disease}, Oxford University Press, 2004.

\bibitem[Rusinek et al.(2004)]{Rusinek_2004}
H. Rusinek, Y. Endo, S. De Santi, D. Frid, W.-H.
Tsui, S. Segal, A. Convit, ``Atrophy rate in medial temporal lobe
during progression of Alzheimer's disease'', {\it Neurology},
vol.63, issue 12, 2354-2359, 2004.

\bibitem[Schott et al.(2005)]{Schott_2005}
J. M. Schott, S. L. Price, C. Frost, J. L. Whitwell, et al,
``Measuring atrophy on Alzheimer disease - A serial MRI study over 6
and 12 months,'' {\it Neurology}, 65(1), pp.119-124, 2005.

\bibitem[Schuff et al.(2010)]{Schuff_2010}
N. Schuff, D. Tosun, P. S. Insel, G. C. Chiang, D. Truran, P. S.
Aisen, C. R. Jack, Jr., M. W. Weiner, the Alzheimer's Disease
Initiative, ``Nonlinear time course of brain volume loss in
cognitively normal and impaired elders'', {\it Neurobiology of
Aging}, 2010, doi:10.1016/j.neurobiolaging.2010.07.012.

\bibitem[Smith et al.(2002)]{FSL-BET}
S. M. Smith, ``Fast robust automated brain extraction,'' {\it Human
Brain Mapping}, 17(3):143-155, 2002.

\bibitem[Shen et al.(2002)]{Shen_HAMMER_ITMI}
D. Shen and C. Davatzikos, ``HAMMER: hierarchical attribute matching
mechanism for elastic registration,'' {\it IEEE Trans. Medical
Imaging}, 21(11):1421-1439, 2002.

\bibitem[Shen et al.(2003)]{Shen_2003}
D. Shen, C. Davatzikos, ``Very High-resolution Morphometry Using
Mass-preserving Deformations and HAMMER Elastic Registration,'' {\it
NeuroImage} 18, pp.28-41, 2003.

\bibitem[SPM website()]{SPM_website}
SPM website: Statistical Parametric Mapping (SPM),
http://www.fil.ion.ucl.ac.uk/spm

\bibitem[Stoub et al.(2005)]{Stoub_2005}
T. R. Stoub, M. Bulgakova, S. Leurgans, D. A. Bennett, D.
Fleischman, D. A. Turner, L. deToledo-Morrell, ``MRI predictors of
risk of incident Alzheimer disease,'' {\it Neurology} 64, pp.
1520-1524, 2005.

\bibitem[Thompson et al.(2003)]{Thompson_2003}
P. M. Thompson, K. M. Hayashi, G. de Zubicaray, A. L. Janke, S. E.
Rose, et al, ``Dynamics of gray matter loss in Alzheimer's
disease,'' {\it J. Neurosci.} 23, pp.994-1005, 2003.

\bibitem[Vapnik(1998)]{Vapnik_1998}
V. Vapnik, {\em Statistical Learning Theory}. John Wiley \& Sons,
1998.

\bibitem[Vemuri et al.(2008)]{Vemuri_2008}
P. Vemuri, J. L. Gunter, M. L. Senjem, J. L. Whitwell, K. Kantarci,
D. S. Knopman, B. F. Boeve, R. C. Petersen, C. R. Jack Jr.,
``Alzheimer's disease diagnosis in individual subjects using
structural MR images: Validation studies,'' {\it NeuroImage} 39,
pp.1186-1197, 2008.

\bibitem[Vemuri et al.(2009)]{Vemuri_2009}
P. Vemuri, H. J. Wiste, S. D. Weigand, L. M. Shaw, J. Q.
Trojanowski, M. W. Weiner, D. S. Knopman, R. C. Petersen, C. R. Jack
Jr., and On behalf of the Alzheimer's Disease Neuroimaging
Initiative, ``MRI and CSF biomarkers in normal, MCI, and AD
subjects: Predicting future clinical change,'' {\it Neurology}
73:294-301, 2009.

\bibitem[Wang et al.(2010)]{Wang_2010}
Y. Wang, Y. Fan, P. Bhatt, C. Davatzikos, ``High-dimensional pattern
regression using machine learning: From medical images to continuous
clinical variables,'' {\it NeuroImage} 50, pp.1519-1535, 2010.

\bibitem[Zhang et al.(2001)]{FSL-FAST}
Y. Zhang, M. Brady, and S. Smith, ``Segmentation of brain MR images
through a hidden Markov random field model and the expectation
maximization algorithm,'' {\it IEEE Trans. Medical Imaging},
20(1):45-57, 2001.

\bibitem[Zhang et al.(2011)]{DZhang_2011}
D. Zhang, Y. Wang, L. Zhou, H. Yuan, D. Shen, and the Alzheimer's
Disease Neuroimaging Initiative, ``Multimodal classification of
Alzheimer's disease and mild cognitive impairment,'' {\it
NeuroImage}, 2011, doi:10.1016/j.neuroimage.2011.01.008.

\end{thebibliography}



\end{document}